\documentclass[12pt]{article}         
\usepackage{amssymb,amsmath,amsbsy,amsfonts,slashed,mathrsfs}
\usepackage{cite}
\usepackage{epsfig}
\usepackage{graphics}
\usepackage{epsf}
\usepackage[pdfborder={0 0 0}, colorlinks=false, breaklinks=false]{hyperref}

\newlength{\dinwidth} 
\newlength{\dinmargin}
\newcommand{\ie}{{\it i.e.\ }}
\newcommand{\eg}{{\it e.g.\ }}
\newcommand{\cf}{{\it cf.\ }}

\newcommand{\nind}{\noindent}

\newcommand{\RR}{\mathbb R}
\newcommand{\CC}{\mathbb C}

\newcommand{\ZZ}{\mathbb Z}

\def\idty{{\leavevmode\hbox{\rm 1\kern -.3em I}}}
\def\net{{\{\As(\Os)\}_{\Os\in\Rs}}}
\def\wnet{{\{\As(\Ws)\}_{\Ws \in \bcW}}}

\def\As{{\cal A}}
\def\Bs{{\cal B}}
\def\Cs{{\cal C}}
\def\Ds{{\cal D}}

\def\Hs{{\cal H}}

\def\Ks{{\cal K}}
\def\Ls{{\cal L}}

\def\Os{{\cal O}}
\def\Ps{{\cal P}}

\def\Rs{{\cal R}}
\def\Ss{{\cal S}}

\def\Ws{{\cal W}}

\def\Pid{{\Ps_+ ^{\uparrow}}}

\def\idty{{\leavevmode\hbox{\rm 1\kern -.3em I}}}

\def\RR{{\mathbb R}}
\def\CC{{\mathbb C}}

\def\IN{{\mathbb N}}
\def\ZZ{{\mathbb Z}}

\def\supp{{\textnormal{supp}}}

\def\CC{{\mathbb C}}

\def\RR{{\mathbb R}}

\def\ZZ{{\mathbb Z}}

\newcommand{\bcW}{{\mbox{\boldmath$\cal W$}}}

\newcommand{\bcp}{{\mbox{\boldmath$p$}}}
\newcommand{\bcq}{{\mbox{\boldmath$q$}}}

\begin{document}

\title{\Large \bf A Perspective on Constructive Quantum Field Theory}

\author{\large { Stephen J.\ Summers\, }\\[4mm]
Department of Mathematics, 
University of Florida, \\ Gainesville FL 32611, USA}

\date{\today}

\maketitle 

\begin{abstract} An overview of the accomplishments of constructive quantum
field theory is provided.\footnote{This is an expanded version of an
article commissioned for UNESCO's Encyclopedia of Life Support Systems 
(EOLSS).}  
\end{abstract}

\section{Introduction: Background and Motivations} \label{intro}

     Quantum field theory (QFT) is widely viewed as one of the most successful
theories in science --- it has predicted phenomena before they were
observed in nature\footnote{For example, the existence and properties of the 
W and Z bosons, as well as the top and charm quarks, were predicted before they
were found experimentally.}, and its predictions are believed 
to be confirmed by experiments to within an extraordinary degree of 
accuracy\footnote{For example, the two parts in one billion difference between 
the theoretical prediction from the Standard Model and the experimentally 
measured value of the anomalous magnetic moment of the muon \cite{DaMa}.}. 
Though it has undergone a long and complex development from its 
origins\footnote{A certain amount of arbitrariness and personal taste must go 
into pointing to a single point of origin, since the 1927 discussion of 
a quantum theory of electromagnetic radiation by Dirac as well as the studies 
of relativistic wave mechanics by Dirac, Schr\"odinger and even de Broglie were 
influential. In any case, the interested reader should see \cite{Schw} for
a detailed account of the birth of QFT.} in the 1929/30 papers of Heisenberg 
and Pauli \cite{HePa1,HePa2} 
and has attained an ever increasing theoretical sophistication, it is still 
not clear in which sense the physically central quantum field theories such as 
quantum electrodynamics (QED), quantum chromodynamics (QCD) and the Standard 
Model (SM) are mathematically well defined theories based upon fundamental
physical principles that go beyond the merely {\it ad hoc}. Needless to say, 
there are many physicists working with quantum field theories for whom the 
question is of little to no interest. But there are also many who are not 
satisfied with the conceptual/mathematical state of quantum field theory and 
have dedicated entire careers to an attempt to attain some clarity in the 
matter.

     This is not the place to explain the grounds for this dissatisfaction;
instead, the goal of this paper is to provide a perspective on ``constructive
quantum field theory'' (CQFT), the subfield of mathematical physics concerned 
with establishing the existence of concrete models of relativistic quantum field
theory in a very precise mathematical sense and then studying their properties
from the point of view of both mathematics and physics. Although the insights 
and techniques won by the constructive quantum field theorists have proven
to be useful also in statistical mechanics and many body physics, these
further successes of CQFT are not discussed here. In addition, we shall restrict
our attention solely to relativistic QFT on $d$ dimensional Minkowski space, 
$d \geq 2$; to this point in time most work in CQFT has been carried out 
precisely in that context. Throughout, as is customary in QFT, we adopt 
physical units in which $c = h/2\pi = 1$.

     In the 1950's and early 1960's various ``axiomatizations'' of QFT were 
formulated. These can be seen to have two primary goals --- (1) to abstract
from heuristic QFT the fundamental principles of QFT and to formulate
them in a mathematically precise framework; (2) on the basis of this framework,
to formulate and solve conceptual and mathematical problems of heuristic QFT
in a mathematically rigorous manner. As it turned out, the study
and further development of these axiom systems led to unanticipated conceptual 
and physical breakthroughs and insights, but these are also not our topic
here.

     The first and most narrow axiomatization scheme of the two briefly 
discussed here is constituted by the {\it Wightman axioms} (see \eg 
\cite{StWi}). This axiom system adheres most closely to heuristic QFT in 
that the basic objects are local, covariant fields acting on a fixed Hilbert 
space. A (scalar Bose) Wightman theory is a quadruple $(\phi, \Hs, U, \Omega)$ 
consisting of a Hilbert space $\Hs$, a strongly continuous unitary 
representation $U$ of the (covering group of the) identity component $\Pid$ 
of the Poincar\'e group acting upon $\Hs$, a unit vector $\Omega \in \Hs$ which 
spans the subspace of all vectors in $\Hs$ left invariant by 
$U(\Pid)$,\footnote{This condition, referred to as the ``uniqueness of the 
vacuum,'' is posited for convenience. With known techniques one can decompose
a given model into submodels that satisfy this condition as well as the 
remaining conditions \cite{Bo,Ar1,DrSu1,DrSu2}.} and an (unbounded) operator 
valued distribution\footnote{Although it is possible, indeed sometimes 
necessary, to choose other test function spaces, here we shall restrict our 
attention to the Schwartz tempered test function space $\Ss(\RR^d)$.} $\phi$ 
such that for every test function $f$, the operator $\phi(f)$ has a dense 
invariant domain $\Ds$ spanned by all products of field operators applied to 
$\Omega$. These conditions are a rigorous formulation of tacit assumptions 
made in nearly all heuristic field theories. In addition, a number of 
fundamental principles were identified and formulated in this framework.

\nind {\bf Relativistic Covariance}: For every Poincar\'e element 
$(\Lambda,a) \in \Pid$ one has \newline
$U(\Lambda,a) \phi(x) U(\Lambda,a)^{-1} = \phi(\Lambda x + a)$,
in the sense of operator valued distributions on $\Ds$.

\nind {\bf Einstein Causality}:\footnote{Also called microscopic causality,
local commutativity or, somewhat misleadingly, locality.} For all spacelike 
separated $x,y \in \RR^4$ one has $\phi(x) \phi(y) = \phi(y) \phi(x)$
in the sense of operator valued distributions on $\Ds$.  

\nind {\bf The Spectrum Condition} (stability of the field system): Restricting
one's attention to the translation subgroup $\RR^4 \subset \Pid$, the
spectrum of the self-adjoint generators of the group $U(\RR^4)$ is contained 
in the closed forward lightcone 
$\overline{V_+} = \{ p = (p_0,p_1,p_2,p_3) \in \RR^4 \mid 
p_0^2 - p_1^2 - p_2^2 - p_3^2 \geq 0 \}$.

\nind The reader is referred to \cite{StWi,Jo} for a discussion of the
physical interpretation and motivation of these conditions. There is an
equivalent formulation of these conditions in terms of the {\it Wightman
functions}  \cite{StWi}
$$W_n(x_1,x_2,\ldots,x_n) \equiv 
\langle \Omega, \phi(x_1) \phi(x_2) \cdots \phi(x_n) \Omega \rangle \, , \, 
n \in \IN \, ,$$
which are distributions on $\Ss(\RR^{dn})$. These two sets of conditions are 
referred to collectively as the {\it Wightman axioms}. There are closely 
related sets of conditions for Fermi fields and higher spin Bose fields 
\cite{StWi,Jo}. 

     A more general axiom system which is conceptually closer to the 
actual operational circumstances of a theory tested by laboratory experiments
is constituted by the {\it Haag--Araki--Kastler axioms} (HAK axioms), which 
axiomatize conceptual structures referred to as local quantum physics or 
algebraic quantum field theory (AQFT). 
Although more general formulations of AQFT are available, for the purposes of 
this paper it will suffice to limit our attention to a quadruple\footnote{In
point of fact, these conditions actually describe an algebraic QFT in a 
(Minkowski space) {\it vacuum representation}. By no means is AQFT limited to 
such circumstances; some other representations of physical interest are
briefly discussed below. Moreover, the algebraic approach to QFT has proven
to be particularly fruitful in addressing conceptual and mathematical problems 
concerning quantum fields on curved spacetimes.} 
$(\net, \Hs, U, \Omega)$ with $\Hs$, $U$ and $\Omega$ as above and
$\net$ a net of von Neumann algebras $\As(\Os)$ acting on $\Hs$, where
$\Os$ ranges through a suitable set $\Rs$ of nonempty open subsets of Minkowski 
space. The algebra $\As(\Os)$ is interpreted as the algebra generated
by all (bounded) observables measurable in the spacetime region $\Os$, so the
net $\net$ is naturally assumed to satisfy isotony: if $\Os_1 \subset \Os_2$, 
then one must have $\As(\Os_1) \subset \As(\Os_2)$.
In this framework the basic principles are formulated as follows.

\nind {\bf Relativistic Covariance}: For every Poincar\'e element 
$(\Lambda,a) \in \Pid$ and spacetime region $\Os \in \Rs$ one has
$U(\Lambda,a) \As(\Os) U(\Lambda,a)^{-1} = \As(\Lambda \Os + a)$.

\nind {\bf Einstein Causality}:\footnote{Also often referred to as locality.} 
For all spacelike separated regions
$\Os_1, \Os_2 \in \Rs$ one has $A B  = B A$ for all $A \in \As(\Os_1)$
and all $B \in \As(\Os_2)$.

\nind {\bf The Spectrum Condition} (stability of the field system): Same as
above.

\nind The reader is referred to \cite{Haag,Ar} for a discussion of the
physical interpretation and motivation of these conditions.
The relation between the Wightman axioms and AQFT is well understood (see
\eg \cite{DSW,BoYn,Bu,FrHe,Su1}). It is important to note that, in general,
infinitely many different fields in the sense of the Wightman axioms
are associated with the same net of observable algebras. Indeed,
an analogy has often been drawn between the choice of a particular coordinate
system, made in order to carry out a computation more conveniently, in 
differential geometry and the choice of a particular field out of the many 
fields associated with a given net. For this and other reasons, those who work 
in mathematical QFT consider nets of observable algebras to be more 
{\it intrinsic} than the associated quantum fields, which are used primarily 
for computational convenience.

     Associated to any Wightman system $(\phi, \Hs, U, \Omega)$ is a net
of *-algebras $\Ps(\Os)$, $\Os \subset \RR^4$. Because all field operators
have the common, dense domain $\Ds \subset \Hs$, arbitrary ``polynomials''
of field operators can be formed on $\Ds$. $\Ps(\Os)$ denotes
the algebra formed by all polynomials (in the sense of functions of infinitely
many variables) in which the supports of all test
functions of all field operators entering into the polynomial are contained
in the spacetime region $\Os$. The algebras $\Ps(\Os)$ are not C*-algebras
but satisfy all of the other HAK axioms. Despite the non-intrinsic nature
of such algebras and despite the technical disadvantages of working with
*-algebras instead of with C*-algebras, mathematical quantum field theorists 
find it convenient for various purposes to work with such nets or with 
similar nets of non-C*-algebras. 

     The goal of constructive QFT, as is it usually understood, is to 
construct in a mathematically rigorous manner physically relevant quantum 
field models which satisfy one of these systems of axioms and then to study 
their mathematical properties with an emphasis on those properties which
can be shown to have physical relevance. This article briefly describes
such models and the means by which they were constructed and is organized both 
historically and by the construction techniques employed.  

     As pointed out independently by Borchers and Uhlmann, the Wightman axioms 
can be understood in a representation independent manner in terms of what is 
now called the Borchers (or Borchers--Uhlmann) algebra --- a tensor algebra 
constructed out of the test function space $\Ss(\RR^4)$ with operations 
directly motivated by the Wightman axioms. Borchers algebras have been 
extensively studied from the point of view of QFT, especially by 
Borchers, Uhlmann, Yngvason and Lassner (see \eg \cite{Ho} for definitions and
references). A Wightman system can be thought of as a concrete representation 
of the Borchers algebra, and for a time there was hope one could arrive at 
quantum field models by defining suitable states on the Borchers algebra and 
employing the standard GNS construction to obtain the corresponding 
representation. However, it proved to be too difficult to conjure such
states.

     The first quantum field models constructed were the free quantum fields,
the Wick powers of such free fields and the so--called generalized free fields. 
These models have been constructed using a variety of techniques (cf. \eg 
\cite{WiGa,Jo,Ar2,Si,GlJa,BSZ,BrGuLo,MuSchYn}) and have been shown to 
satisfy the two axiom systems discussed above; a recent construction of free 
fields which is of particular conceptual interest is briefly described in 
Section \ref{algebra2}. The Hilbert space upon which such fields act is called
the Fock space. Common to these models is the fact that their
S--matrix, the object which describes the scattering behavior of the
``particles'' described by such fields (\cf \cite{Jo,Ar}), is just the 
identity map. 

     We turn now to models with nontrivial S--matrices, \ie 
interacting quantum field models. When referring to the models, we employ the 
standard notation $M_d$, which means quantum model $M$ in $d$ spacetime 
dimensions. Because the mathematical and conceptual difficulties inherent
in the construction of quantum field models are quite daunting, constructive
quantum field theorists proceeded by considering increasingly challenging
models; this often entailed starting the study of the model $M$ with 
$d = 2$, then $d = 3$, and finally $d = 4$. At this point in time only a few 
models have been constructed in four spacetime dimensions. In this respect, the 
reader is referred to Section \ref{outlook} for a few words about the outlook 
for CQFT after more than fifty years of strenuous effort. The reader should 
note 
that all results discussed in this article, unless explicitly stated otherwise, 
are proven according to the criteria accepted by mathematicians and not merely 
on the basis of the plausibility arguments accepted by most physicists as 
``proof''.

\section{\hspace*{-6mm} Algebraic Constructions I} \label{algebra1}
\setcounter{equation}{0}

     Preceded by the 1965 dissertations of Jaffe \cite{Ja0} and Lanford 
\cite{La0}, the first complete constructions of interacting quantum fields were 
carried
out in the late 1960's and early 1970's. In this early work the real time 
models were constructed directly using operator algebras and functional 
analysis as the primary tools. Due to Haag's Theorem, it was known that 
the Hilbert space in which these interacting quantum fields would be defined 
could not be Fock space (see \eg \cite{Su2}). However, because Fock space was 
the sole available starting point at that time, ``cutoffs'' were placed on the 
interacting theories so that they could be realized on Fock space in a 
mathematically meaningful manner. These cutoffs were of two general kinds --- 
finite volume cutoffs and ultraviolet cutoffs --- each addressing 
independent sources of the divergences known in QFT since early in its 
development. Guided by heuristic QFT's division of Lagrangian quantum field 
models into superrenormalizable, renormalizable and nonrenormalizable 
models,\footnote{This classification is based upon the 
perturbation theory associated by Feynman and others with interacting fields, 
viewed as perturbations of free fields.} the constructive 
quantum field theorists began with the simplest category, the 
superrenormalizable models. To be able to address the infinite volume 
divergence without wrestling simultaneously with the ultraviolet divergence, 
constructive quantum field theorists first considered self-interacting bosonic
quantum field models in two spacetime dimensions.

     We begin with Glimm and Jaffe's construction of the 
$(\phi^4)_2$ model \cite{GlJa1,GlJa2,GlJa3,GlJa4}, the self-interacting 
scalar Bose field on two dimensional Minkowski space with Lagrangian 
self-interaction $\lambda \phi^4$, where $\lambda$ is the coupling constant. 
Let $\Hs_0$ be the Fock space for a (free) scalar 
hermitian Bose field $\phi(t,x)$ of mass $m > 0$ ($(t,x) \in \RR^2$). Let 
$\pi(t,x) = \partial \phi(t,x)/\partial t$ be the canonically conjugate 
momentum field and $\Ds\subset\Hs_0$ be the dense set of finite-particle 
vectors in $\Hs_0$. Then, for every $f$ in a dense subspace $\Ss(\RR)$ of 
$L^2(\RR)$, the operator $\phi_0(f) \equiv \int\phi(0,x)f(x) \, dx$ 
is essentially self-adjoint on $\Ds$ and $\phi_0(f)\Ds \subset \Ds$ (similarly 
for $\pi_0(f)$). These operators satisfy the canonical commutation relations 
(CCR) on $\Ds$:
\begin{eqnarray*}
\phi_0(f)\pi_0(g) - \pi_0(g)\phi_0(f) & = & i<f,g>\idty \; , \\
\phi_0(f)\phi_0(g) - \phi_0(g)\phi_0(f) = & 0 & = 
\pi_0(f)\pi_0(g) - \pi_0(g)\pi_0(f) \; ,
\end{eqnarray*}
for all $f,g \in \Ss(\RR)$, where $< \cdot , \cdot >$ is the inner product on
$L^2(\RR)$ and $\idty$ is the identity operator on $\Hs$. When exponentiated 
using the spectral calculus, (the closures of) these operators provide a Weyl 
representation of the CCR. For each bounded open subset $\bf{O} \subset \RR$, 
denote by $\As(\bf{O})$ the von Neumann algebra generated by the Weyl unitaries
\begin{displaymath}
\{e^{i\phi_0(f)},e^{i\pi_0(f)} \mid f \in \Ss(\RR) \; , \; 
\textnormal{supp}(f) \subset \bf{O}\} \; .
\end{displaymath}
($\textnormal{supp}(f)$ denotes the support of the function $f$.)  

     The total energy 
\begin{displaymath}
H_0 = \frac{1}{2}\int \ :(\pi(0,x)^2 + \nabla\phi(0,x)^2 + m^2\phi(0,x)^2): dx
\end{displaymath}
of the free field is a positive quadratic form on $\Ds \times \Ds$ and 
therefore determines uniquely a positive self-adjoint operator, which we also 
denote by $H_0$. The double colons indicate that the expression between them
is Wick ordered, which is a physically motivated way to define in a rigorous
manner a product of operator valued distributions. In this case, the Wick 
ordering is performed with respect to the Fock vacuum (\cf \cite{GlJa7}). 
With $g \in L^2(\RR)$ nonnegative of compact support,  
Glimm and Jaffe showed that, for each $\lambda > 0$, the cut-off interacting 
Hamilton operator
\begin{displaymath}
H(g) \equiv H_0 + \lambda \int :\phi(0,x)^4: g(x) \, dx
\end{displaymath}
is essentially self-adjoint on $\Ds$,\footnote{Without the cutoff $g$, 
the interacting Hamilton operator is \emph{not} densely defined in Fock space.}
and its self-adjoint closure, also denoted by $H(g)$, is bounded from below. 
By adding a suitable multiple of the identity we may take $0$ to be the 
minimum of its spectrum. Then, they proved that $0$ is a simple eigenvalue of 
$H(g)$ with normalized eigenvector $\Omega(g) \in \Hs_0$. 

     For any $t \in \RR$, let ${\bf O}_t$ denote the subset of $\RR$ consisting
of all points with distance less than $\vert t \vert$ to $\bf{O}$. By choosing
the cutoff function $g$ to be equal to $1$ on ${\bf O}_t$, then for any 
$A \in \As({\bf O})$ the operator
\begin{displaymath}
\sigma_t(A) \equiv e^{itH(g)}Ae^{-itH(g)}
\end{displaymath}
is independent of $g$ and is contained in $\As({\bf O}_t)$. For any bounded 
open $\Os \subset \RR^2$ and $t \in \RR$, let
${\bf O}(t) = \{ x \in \RR \mid (t,x) \in \Os \}$ be the time $t$ slice of 
$\Os$. We define $\As(\Os)$ to be the von Neumann algebra generated by
$\bigcup_s \sigma_s(\As({\bf O}(s)))$.\footnote{One can 
then show that the algebra $\As(\Os)$ coincides with the von Neumann algebra 
generated by bounded functions of the self-adjoint field operators 
$\int\phi(t,x)f(t,x)\, dxdt$, with test functions $f(t,x)$ having support in 
$\Os$.}  Finally, we let $\As$ denote the
closure in the operator norm of the union $\bigcup \As(\Os)$ over all
open bounded $\Os\subset\RR^2$. Hence, $\sigma_t$ is an automorphism on $\As$
and implements the time evolution associated with the interacting field. 
Similarly, ``locally correct'' generators for the Lorentz boosts and the 
spatial translations can be defined, resulting in an automorphic action 
$\alpha$ on $\As$ of the entire (identity component of the) Poincar\'e group 
$\Pid$ in two spacetime dimensions.   

     For each $A \in \As$, we set $\omega_g(A) = \;  <\Omega(g), A\Omega(g)>$
($\langle \cdot, \cdot \rangle$ denotes here the inner product on $\Hs$)
to define the locally correct vacuum state $\omega_g$ of the interacting 
field. Taking a limit as the cutoff function $g$ approaches the constant 
function $1$, Glimm and Jaffe showed that $\omega_g(A) \rightarrow \omega(A)$, 
for each $A \in \As$, defines a new (locally normal) state $\omega$ on $\As$ 
which is Poincar\'e invariant, \ie $\omega(\alpha_{(\Lambda,x)}(A)) = \omega(A)$
for all $(\Lambda,x) \in \Pid$ and all $A \in \As$. Employing the GNS 
construction, one then obtains a new Hilbert space $\Hs$, a representation 
$\rho$ of $\As$ as a $C^*$-algebra acting on $\Hs$, and a vector 
$\Omega \in \Hs$ such that $\rho(\As)\Omega$ is dense in $\Hs$ and
\begin{displaymath}
\omega(A) = <\Omega,\rho(A)\Omega> \; , \; \textnormal{for all} \; A \in \As
\; . 
\end{displaymath}
In addition, one obtains a strongly continuous unitary representation $U$ 
of the Poincar\'e group in two spacetime dimensions under which the algebras 
$\rho(\As(\Os))$ transform covariantly, \ie
$$U((\Lambda,x)) \; \rho(\As(\Os)) \; U((\Lambda,x))^{-1} = 
\rho(\As(\Lambda\Os + x))
\, .$$
Both the HAK and Wightman axioms have been verified for this model. 

     The generators of the strongly continuous Abelian unitary groups \newline
$\{ \rho(e^{it\phi(f)}) \mid t \in \RR \}$ and 
$\{ \rho(e^{it\pi(f)}) \mid t \in \RR \}$ satisfy the CCR. However, 
this representation of the CCR in $\Hs$ is not unitarily equivalent to the 
initial representation in Fock space, in accordance with Haag's Theorem. 
Indeed, by taking different values of the coupling constant $\lambda$ in the 
above construction, one obtains an uncountably infinite family of mutually 
inequivalent representations of the CCR (see \cite{Froh}).

     It is in this representation $(\rho,\Hs)$ that the field equations for 
this model find a mathematically satisfactory interpretation, as shown by
Schrader \cite{Sch3}. And it is to the physically significant quantities in 
this representation that the corresponding perturbation series in $\lambda$ is 
asymptotic --- see below for further discussion. For this and other reasons, 
$\omega$ is interpreted as the exact vacuum state in the interacting theory 
corresponding to the Lagrangian interaction $\lambda \phi^4$, and the folium 
of states associated with this representation contains the physically 
admissible states of the interacting theory. Many further properties of 
physical relevance have been proven for this model more recently --- 
see the discussion below in Section \ref{P}.

     The results attained for the $\phi^4_2$ model were subsequently extended 
to $P(\phi)_2$ models (using a periodic box cutoff), where $P(\phi)$ is any 
polynomial bounded from below \cite{GlJa9,GlJa10,GlJa11}.\footnote{If $P(\phi)$
is not bounded from below, then the corresponding cutoff Hamiltonian $H(g)$
is not bounded from below and the resulting model is not stable.} (See
\cite{GlJa11,GlJa7} for more complete references and history of this
development.) Hoegh-Krohn employed the techniques of Glimm and Jaffe
to construct models in two spacetime dimensions (with similar results) in which 
the polynomial interaction $P(\phi)$ is replaced by a function of exponential 
type, the simplest example being $e^{\alpha \phi}$ \cite{HK1}.

     Analogous results were proven for Y$_2$, the Yukawa model in two
spacetime dimensions by Glimm and Jaffe and Schrader \cite{GlJa5,GlJa6,Sch2}. 
In this model one commences with the direct product 
$\Hs_0 = \Hs_b \otimes \Hs_f$ of the Fock space $\Hs_b$
for a scalar hermitian Bose field $\phi(t,x)$ of mass $m_b > 0$ and the
Fock space $\Hs_f$ for a Fermi field $\psi(t,x)$ of mass $m_f > 0$.
In this model the free Hamiltonian $H_0$ is the total energy operator of
the free fields $\phi$ and $\psi$. Because there is still an ultraviolet
divergence remaining after Wick ordering, the cutoff interacting Hamiltonian
is $H(g,\kappa) \equiv H_0 + H_I(g,\kappa) + c(g,\kappa)$, where 
$H_I(g,\kappa)$ is the result of applying a certain multiplicative ultraviolet
cutoff (which is removed in the limit $\kappa \rightarrow \infty$) to the 
formal expression  
$$H_I(g) \equiv \lambda \int g(x) \; \phi(0,x) \; 
:\overline{\psi} \psi :(0,x)\,  dx \, ,$$
and $c(g,\kappa)$ is a (finite) renormalization counterterm determined by 
second-order perturbation theory which diverges as $\kappa \rightarrow \infty$
\cite{GlJa5} and includes both a mass and vacuum energy renormalization. With 
both volume and ultraviolet cutoffs in place, $H(g,\kappa)$ is a well defined 
operator on $\Hs_0$. Glimm and Jaffe show that as $\kappa \rightarrow \infty$
the operator $H(g,\kappa)$ converges in the sense of graphs to a positive
self-adjoint operator $H(g)$ with an eigenvector $\Omega(g) \in \Hs_0$ of 
lowest energy 0. Once again, they control the limit as $g \rightarrow 1$ of 
the expectations $\omega_g(A)$ for all $A \in \As_b \otimes \As_f$ and obtain 
a state $\omega$ on $\As_b \otimes \As_f$ that provides a corresponding (GNS) 
representation of the fully interacting theory. Glimm and Jaffe  \cite{GlJa6}
also proved that the Yukawa field equations are satisfied by the fields
in that representation. A similar argument was applied to the
Y$_2$ + $P(\phi)_2$ model by Schrader \cite{Sch1}), where 
$$H_I(g) \equiv \lambda \int g(x) \; (\phi(0,x) \; 
:\overline{\psi} \psi :(0,x) \, + \, : P(\phi):(0,x) ) \; dx $$
and $P(\phi)$ is any polynomial bounded from below. The axioms of HAK and 
Wightman were shown to hold in these models (see \cite{Su0}), at least 
for all sufficiently small values of the coupling constant $\lambda$. In 
addition, by using a mixture of algebraic and Euclidean methods Summers
\cite{Su0} showed that the model manifests further properties of physical 
relevance, such as the funnel property (also known as the split property) and 
all assumptions of the Doplicher--Haag--Roberts superselection theory 
(\cf \cite{Ar,Haag}). Therefore the model also admits the physically 
expected Poincar\'e covariant, positive energy, charged representations 
associated with the generator of the global U(1) gauge group of the model, 
which are mutually unitarily inequivalent.

     Along the lines employed in the construction of the Yukawa model in two 
spacetime dimensions, Glimm and Jaffe \cite{GlJa8} also showed for the 
$\phi^4_3$ model that the spatially cutoff Hamiltonian $H(g)$ is densely 
defined, symmetric and bounded below by a constant $E(g)$ proportional to the 
volume of the support of $g$. The renormalization constants in the Hamiltonian 
$H(g)$ are again given by perturbation theory and involve counterterms to the 
vacuum energy and the rest mass of a single particle. The proof was technically 
more challenging than that for Y$_2$, even though the results were more limited 
to a significant extent. There was real motivation to find an alternative 
approach, as described in the next section.

     However, before proceeding to the next section we mention the
Federbush model, a model of self-interacting fermions in two spacetime 
dimensions. First proposed by Federbush \cite{Fe1}, the Lagrangian of
the model is
$$\sum_{s = \pm 1} \overline{\psi}_s(\slashed\partial - m(s))\psi_s -
2\pi \lambda \epsilon_{\mu\nu} J_1^\mu J_{-1}^\nu \, , $$
where $\epsilon_{10} = -\epsilon_{01} = 1$, $\epsilon_{00} = \epsilon_{11} = 0$,
$J_s^\mu = \overline{\psi}_s \gamma^\mu \psi_s$ and $m(s) > 0$, $s = \pm 1$.
Without cutoffs of any kind, a concrete realization of the Federbush model
can be given in terms of certain exponential expressions on a suitable
Fock space, and Ruijsenaars \cite{Ru1} proved that this realization
satisfies the Wightman axioms when $\lambda \in (-\frac{1}{2},\frac{1}{2})$
(Einstein causality is actually verified only for sufficiently small $\lambda$).
Of particular interest, he proved that the associated Haag--Ruelle
scattering theory is asymptotically complete. The S--matrix is nontrivial,
but there is no particle production \cite{Ru2}. The Federbush model was the
first non-superrenormalizable model for which any of these properties have
been proven.

\section{Functional Integral Constructions --- Euclidean}  \label{path1}
\setcounter{equation}{0}

   The technical difficulties of the approach described in the preceding
section were formidable, and in the early 1970's a technically more
manageable program replaced (at least temporarily) the earlier 
approach.\footnote{See \cite{GlJa,Si,Sy1,Sy2} for some history and references.}
In this newer program a (scalar Bose) Euclidean quantum field is a random 
variable valued distribution $\phi(x)$ with probability measure $\mu$ on the 
corresponding distribution space. The moments of this measure
$$S_n(x_1,x_2,\ldots,x_n) \equiv 
\int \; \phi(x_1) \phi(x_2) \cdots \phi(x_n) \; d\mu(\phi) \, , \, 
n \in \IN \, ,$$
are called the {\it Schwinger functions}. Formally, the connection between
the Wightman functions and the Schwinger functions is described as follows:
Let $x = (x_0,x_1,\ldots,x_{d-1})$ be the coordinates of a point in $d$
dimensional Minkowski space and consider $x^E = (-ix_0,x_1,\ldots,x_{d-1})$.
Then   
$$S_n(x_1,x_2,\ldots,x_n) = W_n(x^E_1,x^E_2,\ldots,x^E_n) \, , \, 
n \in \IN \, .$$
In point of fact, it is known that the Wightman functions of a Wightman
theory can be analytically continued to such Euclidean points
$(x^E_1,x^E_2,\ldots,x^E_n)$ \cite{Jo,Sy1} so the above relation is
meaningful in that sense. Under certain additional
conditions the resulting Schwinger functions are the moments of a Euclidean 
invariant probability measure $\mu$. For example, the Schwinger 
functions of the free scalar Bose field of mass $m_0 > 0$ on two dimensional 
Minkowski space are the moments of the Gaussian probability measure $\mu_C$ on 
$\Ss'(\RR^2)$ with mean 0 and covariance $C = (-\triangle + m_0^2)^{-1}$.
The corresponding two-point Schwinger function is
$$S_2(x - y) = (2\pi)^{-2} \, \int \; \frac{e^{ik \cdot (x - y)}}{k^2 + m_0^2} \,
dk^2 \, , $$
where all inner products are Euclidean. 

     Hence, from a Wightman field theory one can obtain in this manner
a Euclidean field theory. Under certain circumstances one can also
obtain a Wightman field theory from a Euclidean field theory by appropriate
analytic continuation of Schwinger functions to obtain Wightman functions
satisfying the Wightman axioms. Of the various sets of conditions known
to permit this reconstruction of the real time theory from the imaginary
time theory, we mention only the {\it Osterwalder--Schrader axioms} \cite{OS}.
Here one commences with a sequence $\{ S_n(x_1,x_2,\ldots,x_n) \}_{n \in \IN}$
of tempered distributions (again called Schwinger functions) completely 
symmetric in their arguments which are invariant under the natural action of 
the Euclidean group. One requires additionally of these distributions that 
they satisfy a certain cluster property (ergodicity under the ``time'' 
translations), the Osterwalder--Schrader positivity condition (also called 
reflection positivity) and a technical linear growth condition. Osterwalder 
and Schrader 
showed that an inductive construction of analytic continuations applied to 
these Schwinger functions yields a set of Wightman functions satisfying the 
Wightman axioms. They also proved that analytically continuing the Wightman 
functions of a Wightman theory in the manner indicated above results in 
Schwinger functions satisfying all of the Osterwalder--Schrader axioms except 
possibly the linear growth condition. Since the linear growth condition may 
fail, the Osterwalder--Schrader axioms may not be equivalent to the Wightman 
axioms.\footnote{Finding a physically meaningful set of conditions on 
Schwinger functions which is equivalent to the Wightman axioms is an open 
problem --- see \cite{Schm} for a recent overview of the topic ``equivalence
of Wightman functions and Schwinger functions'' with references.}

     In general, a given sequence of Schwinger functions satisfying the
Osterwalder--Schrader axioms need not be the moments of a measure $\mu$.
However, the general strategy of the Euclidean construction program was to
construct a suitable probability measure $\mu$ on $\Ss'(\RR^d)$ such that
its moments satisfy the Osterwalder--Schrader axioms. Application of the 
Osterwalder--Schrader reconstruction theorem would then result in a
corresponding relativistic quantum field model satisfying the Wightman
axioms. One would then further study these models for properties of physical
relevance that go beyond the basic axioms. This approach also stimulated
a fruitful exchange of techniques and tools between quantum field theory
on one side and probability theory and classical statistical mechanics on
the other. 

     The connections between Euclidean QFT and probability theory are 
even richer than what has been suggested above. Nelson \cite{Ne} pointed
out that the free scalar massive Bose Euclidean field is a Markov process,
as is the finite volume $P(\phi)_2$ field. This line of thought led to
the use of the stochastic processes which are solutions of suitable
stochastic differential equations to construct models of quantum fields
--- see Section \ref{stoch}.

\subsection{$P(\phi)_2$ models} \label{P}

     The $P(\phi)_2$ models were reexamined from the point of view of the
Euclidean construction program, and new results of physical relevance were 
obtained. In analogy to the real time construction sketched in Section 
\ref{algebra1}, one commences with a local perturbation of the free field, 
here represented by the Gaussian measure $\mu_C$. Once again, let $P(\phi)$ 
be a polynomial bounded from below and $\Lambda$ be a rectangular region in 
Euclidean $\RR^2$. The Wick ordered powers 
of the Euclidean free field may be defined recursively by the conditions 
$:\phi^0: = 1$, $\partial :\phi^n: / \partial \phi = n : \phi^{n-1} :$ and  
$\int \; :\phi^n: \, d\mu_C = 0$, $n \in \IN$. Let 
$U_\Lambda \equiv \int_\Lambda \; :P(\phi):(x) \; d^2x$. Then 
$e^{-U_\Lambda} \in L^p(d\mu_C)$ for all $p < \infty$ \cite{Si}, so 
\begin{equation} \label{pp}
\mu_\Lambda \equiv \frac{e^{-\lambda U_\Lambda} \mu_C}{\int e^{-\lambda U_\Lambda} d\mu_C} 
\end{equation}
is a well defined probability measure on $\Ss'(\RR^2)$, where the coupling
constant $\lambda \geq 0$. One may then consider the corresponding Schwinger 
functions
$$S_n^{\Lambda}(x_1,x_2,\ldots,x_n) \equiv 
\int \; \phi(x_1) \phi(x_2) \cdots \phi(x_n) \; d\mu_\Lambda \, , \, n \in \IN 
\, . $$
By employing cluster expansions Glimm, Jaffe and Spencer proved that 
the limits
$$S_n(x_1,x_2,\ldots,x_n) \equiv \lim_{\Lambda \nearrow \RR^2} 
S_n^{\Lambda}(x_1,x_2,\ldots,x_n) \, , \, n \in \IN $$
exist\footnote{In point of fact, the measures $\mu_\Lambda$ converge weakly to
a measure $\mu_\infty$ whose moments are $S_n(x_1,x_2,\ldots,x_n)$, $n \in \IN$.} 
and satisfy the Osterwalder--Schrader axioms for all sufficiently
small $\lambda / m_0^2$ \cite{GJS}. Hence, one obtains a corresponding
relativistic quantum field model.

     Similar results can also be obtained, at least for a large class of
polynomials, by commencing with a lattice approximation and controlling
the limit as the lattice spacing goes to 0 \cite{GlJa,Si} and/or by using 
correlation inequalities instead of cluster expansions \cite{GlJa,Si}. 
In the former, placing the quantum field model on a lattice results in
a classical Ising ferromagnetic system with unbounded spins and nearest neighbor
coupling arising from the finite difference approximation to $-\Delta + m_0^2$. 
The lattice spacing serves as an ultraviolet cutoff. An example of the 
latter are the GKS-inequalities, which were extended from lattice ferromagnetic
systems to $P(\phi)_2$ models by Guerra, Rosen and Simon \cite{GuRoSi}: For 
$P(\phi) = Q(\phi) - h \phi$, where $Q$ is an even polynomial and $h \geq 0$,
one has
$$ S_n^{\Lambda}(x_1,x_2,\ldots,x_n) \geq 0 \, , \, n \in \IN \, ,$$
and 
$$S_{n+m}^{\Lambda}(x_1,x_2,\ldots,x_{n+m}) \geq S_n^{\Lambda}(x_1,x_2,\ldots,x_n)
S_m^{\Lambda}(x_{n+1},x_{n+2},\ldots,x_{n+m}) \, , \, n,m \in \IN \, ,$$
in the sense of distributions. Nelson proved that the Schwinger functions are 
monotone in the volume cutoff $\Lambda$,\footnote{Strictly speaking, 
``half-Dirichlet boundary conditions'' on $\Lambda$ are introduced to prove 
this.} namely for suitably regular regions $\Lambda^\prime \subset \Lambda$ one 
has
$$ S_n^{\Lambda^\prime}(f_1,f_2,\ldots,f_n) \leq S_n^{\Lambda}(f_1,f_2,\ldots,f_n) $$
for all positive test functions $f_1,\ldots,f_n$ and all $n \in \IN$ \cite{Ne}.
Bounds on the Schwinger functions which are uniform in $\Lambda$ yield the
existence of the infinite volume limit. The GKS inequalities are used in the
proof of both the monotonicity and the uniform bounds. Further arguments yield 
the validity of the Osterwalder--Schrader axioms for the limit Schwinger 
functions. Though use of correlation inequalities restricts the interaction 
polynomial, wherever correlation inequalities can be used, there is no 
restriction on the size of $\lambda \geq 0$. This is an advantage 
with respect to the results attained using cluster expansions.

     In the weak coupling limit (sufficiently small $\lambda / m_0^2$) the 
``uniqueness of the vacuum'' follows from exponential clustering bounds
on the Schwinger functions (or the Wightman functions): there exists an $m > 0$
such that for all $0 \leq k \leq n \in \IN$ and $a \in \RR^2$
\begin{eqnarray}  \label{mass}
S_n(x_1,x_2,\ldots,x_k,x_{k+1} - a,x_{k+2} - a,\ldots,x_n - a) & - &  \\
S_k(x_1,x_2,\ldots,x_k) \; S_{n-k}(x_{k+1},x_{k+2},\ldots,x_n) & \leq & 
Ce^{-m \Vert a \Vert}
\, . \end{eqnarray}
In fact, the bound (\ref{mass}) establishes also the existence of a mass gap
at least as large as $m$, \ie the spectrum of the mass operator is contained in
$\{ 0 \} \cup [m,\infty)$. Indeed, it was shown that there exists an $m > 0$
so that the spectrum of the mass operator is contained in 
$\{ 0,m \} \cup [m_1,\infty)$ and that as $\lambda \rightarrow 0$ with
$m_0$ fixed, one has $m \rightarrow m_0$ and $m_1 \rightarrow 2m_0$ \cite{GJS}. 
Hence, weakly coupled $P(\phi)_2$ models have an isolated one-particle 
hyperboloid, and the Haag--Ruelle scattering theory (see \cite{Jo,Ar}) may be
applied \cite{GJS}, resulting in a well defined scattering theory for these 
models. It has been shown by Osterwalder and S\'en\'eor \cite{OsSe} that the 
resultant S--matrix is nontrivial and that the usual perturbation expansion 
in the coupling constant for this S--matrix is asymptotic to the exact 
S--matrix. In these models there is particle production.

     Much work has been expended in the study of the particle structure
of $P(\phi)_2$ models which goes beyond the existence of an isolated 
mass hyperboloid, since this is of direct physical interest and also
because it appears likely that such knowledge would be a necessary
prerequisite for any proof of the asymptotic completeness of the models.
An early example is the proof by Spencer and Zirilli \cite{SpZi} that
in the $\lambda \phi^4_2$ model there are no even bound states of energy
less than $4(m - \epsilon)$, where $m = m(\lambda)$ is the physical mass of 
the model and $\epsilon \rightarrow 0$ as $\lambda \rightarrow 0$. On the 
other hand, Dimock and Eckmann \cite{DiEc} showed that in the 
$\lambda (\phi^6 - \phi^4)_2$ model there exists a unique
two particle bound state for sufficiently small $\lambda$, and it has mass 
$2m(1 - \frac{9}{8m^4} \lambda^2 + O(\lambda^3))$. They subsequently
generalized these results \cite{DiEc2} to any polynomial interaction having
no positive $\phi^4$ term. They further showed that if there is such a 
$\phi^4$ term, then there is no two particle bound state. In both cases, there 
are no bound states masses embedded in the continuum below $3m - \epsilon$, 
\ie the model is two-particle asymptotically complete. Neves da Silva 
\cite{Nev} proved that when the interaction polynomial is even and there exists 
a two particle bound state, then there also exists a three particle bound 
state of energy less than $3m$. The uniqueness of the former entails the
uniqueness of the latter. Further details of particle structure are strongly 
model dependent, but general results attained by Bros and Iagolnitzer 
\cite{BrIa} are a useful starting point. 

     Using Euclidean Markov fields and correlation inequalities, Albeverio
and Hoegh-Krohn \cite{AlHK} proved that the infinite volume limit of scalar 
bosons with even exponential self-interaction in two space-time dimensions
satisfies the Osterwalder--Schrader axioms, including clustering, so that the 
corresponding Minkowski space theory satisfies the Wightman axioms, including 
uniqueness of the vacuum. Together with Gallavotti they showed \cite{AlGaHK} 
that when $d \geq 3$ the model can also be constructed, but its
Schwinger functions coincide with those of the corresponding free field,
\ie it is trivial.

     In general QFT the ``uniqueness of the vacuum'' can fail in a 
physically significant manner --- the quantum field model can manifest 
``phase transitions'' entirely analogous to those found in statistical
mechanics. In this circumstance the subspace $\Hs_0$ of vectors in $\Hs$ 
invariant under the action of $U(\Pid)$ is not one-dimensional, and one
can find two unit vectors $\Omega_1,\Omega_2 \in \Hs_0$ and an observable
$A$ (often called the order parameter) such that 
$\langle \Omega_1, A \Omega_1 \rangle \neq \langle \Omega_2, A \Omega_2 \rangle$.
As in statistical mechanics, one can determine ``phase diagrams'' plotted
against the various parameters of interaction in the model with ``critical
points'' where phase transition curves end. 

     The first such results were proven for 
$P(\phi) = (\phi^2 - \sigma^2)^2/\sigma^2$ for $\sigma$ a sufficiently large
constant. By a suitable change of variables and Wick ordering, this is 
physically equivalent to the choice 
$P(\phi) = \lambda \phi^4 + \frac{1}{2} m_0^2 \phi^2$ for 
sufficiently small $m_0^2/\lambda$.\footnote{In direct analogy to statistical 
mechanics, the limit $m_0^2/\lambda \rightarrow 0$ is referred to as the low 
temperature limit and that of $\lambda/m_0^2 \rightarrow 0$ is called the high 
temperature limit.} As in statistical mechanics, one displays the phase
transition by considering the interaction 
$P(\phi) = (\phi^2 - \sigma^2)^2/\sigma^2 - h \phi$, with $h$ a constant 
representing an ``external field''. For $h \neq 0$ the corresponding model
has a unique vacuum, by the Lee-Yang theorem generalized to the $\phi^4_2$
model by Simon and Griffiths \cite{SiGr}, \ie the corresponding measure 
$\mu_{\sigma,h}$ is ergodic. For all sufficiently large $\sigma$, Glimm, Jaffe 
and Spencer show that 
$$\lim_{h \searrow 0} \; \langle \phi(f) \rangle_{\sigma,h} = 
- \, \lim_{h \nearrow 0} \; \langle \phi(f) \rangle_{\sigma,h} > 0 \, ,$$
for positive test function $f$, where $\langle \cdot \rangle_{\sigma,h}$
is the expectation with respect to $\mu_{\sigma,h}$. Hence, there is a phase
transition at $h = 0$ for sufficiently large $\sigma$ and the field itself
is an order parameter. Indeed, Glimm, Jaffe and Spencer \cite{GJS1} 
subsequently modified the cluster expansion to show that with 
$P(\phi) = \lambda \phi^4 - \frac{1}{4} \phi^2 - h \phi - E_c$ ($E_c$ chosen so 
that the infimum of the polynomial is 0) and $\mu_b$ the Gaussian measure
on $\Ss^\prime(\RR^2)$ with mean $b = \pm (8\lambda)^{-1/2}$ (note these are the 
minima of $P(\phi)$) and covariance $C = (-\Delta + 1)^{-1}$, then if $h$ and 
$b$ have the same sign the Schwinger functions associated with the measure
\begin{equation} \label{phase}
\mu_\Lambda \equiv 
\frac{e^{-\int_\Lambda \, (:P(\phi):(x) + \frac{1}{2}:(\phi - b)^2:(x)) d^2x} \; d\mu_b}{
\int e^{-\int_\Lambda \, (:P(\phi):(x) + \frac{1}{2}:(\phi - b)^2:(x)) d^2x} \; d\mu_b}
\end{equation}
converge as $\Lambda \nearrow \RR^2$ to Schwinger functions satisfying the
Osterwalder--Schrader axioms for all sufficiently small $\lambda$, in
particular, the corresponding vacuum is unique and the limit measure $\mu_\pm$
is ergodic. Note that in (\ref{phase}) the mean and mass in the Gaussian 
measure exactly cancel the term added to $P(\phi)$ in the region $\Lambda$, 
leaving a $\pm$ boundary condition outside of $\Lambda$. What is more,
$\langle \phi(x) \rangle_\pm = b + O(\lambda^{3/2})$, where 
$\langle \cdot \rangle_\pm$ is the expectation in $\mu_\pm$. Hence, for
$h = 0$ one obtains two distinct Wightman theories in which the corresponding
vacuum expectations of the field differ by a sign and in which the 
$\phi \rightarrow -\phi$ symmetry of the action is spontaneously broken.
In analogy to situations arising in statistical mechanics, one says that at 
$h = 0$ there are two distinct ``phases'' for the model. Koch \cite{Ko2}
showed that in each of these pure phases the mass hyperboloid is isolated,
so that Haag--Ruelle scattering theory may be applied, and that there exists
a two particle bound state.

     The mean field cluster expansion of Glimm, Jaffe and Spencer was 
subsequently extended by Summers \cite{Su00} to an 
interaction involving a sixth degree polynomial, which resulted in a value of 
the interaction coefficients at which three distinct phases coexist and in the 
corresponding phase diagram there are phase lines which are not associated 
with spontaneous symmetry breaking. There are also values of the coefficients 
resulting in two ``critical points'' where phase lines simply end, \ie where 
the distinction between the phases vanishes. Each of the pure phases satisfies 
all of the Osterwalder--Schrader axioms, and the perturbation series for the 
Schwinger functions (suitably adjusted about the mean value of the field in 
the phase) is asymptotic to the exact (similarly adjusted) Schwinger functions. 
Imbrie \cite{Im} generalized this study of phase transitions in $P(\phi)_2$ 
models to very general polynomials yielding quite complex phase diagrams by
suitably adapting techniques developed in statistical mechanics to this
situation. 

     Although most effort has been expended on the construction of quantum
field models in a vacuum representation, other representations are of
physical interest, as well. One such class of representations studied by 
constructive and axiomatic quantum field theorists is the class of equilibrium 
thermal representations, \ie representations associated with states which 
satisfy the KMS condition for some fixed inverse temperature (\cf 
\cite{BrRo} for further background and references). Here we only mention 
results proven in concrete interacting models. Hoegh-Krohn \cite{HK2} 
considered scalar Bose fields in two spacetime dimensions with either 
polynomial or exponential self-interaction and showed using a hybrid of 
algebraic and Euclidean techniques that for any $T > 0$ the 
infinite volume limit of the spacetime cutoff Gibbs state at temperature $T$ 
exists on a suitable global algebra of observables and satisfies both the KMS 
condition for $\beta = T^{-1}$ and the cluster property and is translation
invariant. More recently, G\'erard and J\"akel \cite{GeJa1} revisited
the matter and provided a different proof of these results, as well as
some further, more technical observations.
 
     Thermal equilibrium states cannot be Lorentz invariant \cite{Oj}, even if
the equations of motion are invariant under Poincar\'e transformations and
the signal propagation speed is finite. Nonetheless, as recognized by Bros 
and Buchholz \cite{BrBu}, the {\it passivity} (\cf \cite{BrRo}) of an
equilibrium state should still be visible to an observer in motion with
respect to the rest frame distinguished by the KMS state. They showed that
this entails that the thermal equilibrium states of a relativistic QFT should 
have stronger analyticity properties in configuration space than those already 
implicit in the KMS condition. They formulated
these properties as a {\sl relativistic KMS condition}. Using the Euclidean
approach to thermal fields and technical advances in treating spatially cutoff
models due to Klein and Landau \cite{KlLa}, G\'erard and J\"akel \cite{GeJa2} 
proved that the two-point function in the $P(\phi)_2$ model satisfies this 
relativistic KMS condition, and J\"akel and Robl \cite{JaRo} extended the 
result to general $n$-point functions in this model.

     Other non-vacuum representations of physical interest which have been
studied rigorously in concrete models are {\it charged} representations, an
example of which is provided by the Yukawa$_2$ model \cite{Su0}, where
the countably infinite eigenspaces of a global ``charge'' operator $Q$ 
commuting with all observables and the representation of the Poincar\'e group 
yield mutually inequivalent representations of the observables 
(``superselection sectors'') that satisfy the HAK axioms, excepting the 
existence of the vacuum (the vacuum sector is the charge 0 eigenspace
of $Q$). 

    {\it Soliton} representations can arise with the appearance of 
phase transitions and associated spontaneous breaking of 
symmetries of the model, as in the $\phi^4_2$ models discussed above. They
are believed to occur primarily in models in two spacetime dimensions,
though there are heuristic indications that they might exist in nonabelian 
Yang--Mills models in four spacetime dimensions. They are Poincar\'e covariant,
positive energy representations of the fields (observables) that, in a certain 
sense, interpolate between two distinct vacuum representations. As an 
illustration, consider the $\phi^4_2$ model in the parameter range where the 
phase transition discussed above occurs. At $h = 0$ there exist two vacuum 
vectors $\Omega_\pm$ inducing two vacuum states 
$\omega_\pm( \cdot ) \equiv \langle \Omega_\pm , \cdot \Omega_\pm \rangle$
on the fields, equivalently on the algebra of observables $\As$. As
shown by Fr\"ohlich \cite{Froh3}, there exists a localized automorphism 
$\sigma$ on $\As$ such that $\omega_\pm \circ \sigma$  are states on $\As$
and the GNS representations of $\As$ corresponding to 
$\omega_\pm \circ \sigma$ are Poincar\'e covariant and satisfy the spectrum
condition. If $\Hs_\pm$ are the representation spaces of $\As$ corresponding
to $\omega_\pm$ and $\Hs_{s,s'}$ the representation spaces corresponding to
$\omega_\pm \circ \sigma$, then $\Hs_+ \oplus \Hs_- \oplus \Hs_s \oplus \Hs_{s'}$
is the full Hilbert space of the $\phi^4_2$ model (in the stated parameter
range). On this space a time-independent ``topological'' charge 
operator is defined formally by
$$Q = \int \, \frac{\partial}{\partial x} \phi(t,x) \; dx \, .$$
$\Hs+ \oplus \Hs_-$ is the eigenspace of $Q$ corresponding to the eigenvalue
0, and $\Hs_s$, resp. $\Hs_{s'}$, is the eigenspace corresponding to the 
eigenvalue $2 \langle \Omega_+, \phi(x) \Omega_+ \rangle > 0$, resp.
$2 \langle \Omega_-, \phi(x) \Omega_- \rangle = 
- 2 \langle \Omega_+, \phi(x) \Omega_+ \rangle$. The sense in which the 
soliton, resp. anti-soliton, state $\omega_s$, resp. $\omega_{s'}$, interpolates
between the two vacua is suggested by 
$$\lim_{x \rightarrow \pm \infty} \omega_s(\phi(t,x)) = \omega_\pm(\phi(t,x))$$
and
$$\lim_{x \rightarrow \pm \infty} \omega_{s'}(\phi(t,x)) = \omega_\mp(\phi(t,x)) \, .$$
Although not all details have been published, there are strong indications
in the work of Fr\"ohlich and Marchetti \cite{Froh3,FrMa} that there are
single soliton (antisoliton) states in $\Hs_s$ ($\Hs_{s'}$) which are created
out of the vacuum vectors $\Omega_+$ ($\Omega_-$) by a local field $s(x)$, 
called the soliton field. B\'ellisard, Fr\"ohlich and Gidas \cite{BeFrGi} have 
proven that for small enough coupling $\lambda$ the mass of the soliton grows 
like $\lambda^{-1}$.

     These ideas have been explored also in $P(\phi)_2$ models, the Yukawa
model and others. Of particular interest is the sine--Gordon model
(see Section \ref{sG}), which, because of the periodicity of the interaction,
admits countably infinitely many soliton sectors \cite{Froh3}. These and other 
models have also been studied from the Euclidean point of view by Fr\"ohlich 
and Marchetti \cite{FrMa}, casting a new perspective on these results. See
also Schlingemann \cite{Schl} for more recent developments.

\subsection{The $\phi^4_3$ model}  \label{4}

     Also the $\phi^4_3$ model was revisited using Euclidean techniques,
yielding stronger results. As seen in Section \ref{algebra1}, the $\phi^4_3$ 
model is superrenormalizable and requires an infinite mass renormalization.  
Hence, both a volume and an ultraviolet cutoff are necessary to begin with
well defined quantities. Let $\mu_{C_\kappa}$ be the Gaussian measure on
the Schwartz distribution space $\Ss^\prime(\RR^3)$ with mean 0 and
covariance $C_\kappa$, with kernel 
\begin{equation} \label{cutoff}
C_\kappa(x - y) = (2\pi)^{-3} \, \int \; \frac{e^{ik \cdot (x - y)}}{k^2 + m_0^2} 
\, \eta(k/\kappa) \, dk^3 \, , 
\end{equation}
$m_0 > 0$, where $\eta$ is a smooth function of compact support taking 
the value 1 in a neighborhood of the origin in $\RR^3$. The cutoff free field
$\phi_\kappa$ is the corresponding random variable valued distribution. 
The cutoff action is
$$V(\lambda,\Lambda,\kappa) \equiv \int_\Lambda \, \lambda :\phi_\kappa^4:(x) + 
c(\kappa,\lambda,\phi_\kappa(x)) \; d^3 x \, ,$$
where $c(\kappa,\lambda,\phi_\kappa(x))$ is again a renormalization counterterm 
suggested by perturbation theory (see \eg \cite{MaSe1}), and $\lambda > 0$ is 
the coupling constant. Define the cutoff interacting measure to be 
$$\mu_{\lambda,\Lambda,\kappa} \equiv  \frac{e^{ -V(\lambda,\Lambda,\kappa)} \, \mu_{C_\kappa}}
{\int \, e^{ -V(\lambda,\Lambda,\kappa)} \, \mu_{C_\kappa}} \, . $$
Using phase cell and cluster expansions, Feldman and Osterwalder \cite{FeOs}, 
on one hand, and Magnen and S\'en\'eor \cite{MaSe1}, on the other, have 
independently shown that for sufficiently small $\lambda/m_0^2$ the 
corresponding Schwinger functions converge as $\kappa \rightarrow \infty$, 
$\Lambda \nearrow \RR^3$, to Schwinger functions satisfying the 
Osterwalder--Schrader axioms, so that there is a corresponding Wightman theory. 
By using different techniques going back to Nelson and Guerra, Seiler and 
Simon \cite{SeSi} were able to remove the restriction on $\lambda/m_0^2$. 
Using a lattice approximation and correlation inequalities, Park \cite{Pa} 
attains the same results. The resulting quantum fields
satisfy the corresponding $\phi^4$ field equations \cite{FeRa}, are locally
associated with a net of algebras satisfying the HAK axioms \cite{SeSi},
and the perturbation expansion for the Schwinger functions is Borel
summable \cite{MaSe3}. Burnap \cite{Bur} showed the existence of one particle 
states in this model, so the Haag--Ruelle scattering theory is applicable to
this model.
     
     In the intervening thirty years the model has been revisited from many 
points of view; much effort has been exerted to simplify the techniques used 
in the original proofs (\cf the recent \cite{MaRi}), and we mention, in 
particular, that various rigorous versions of the renormalization group have 
been developed and applied to the $\phi^4_3$ model (cf. \cite{BrDiHu} for an 
overview). This approach and a second, using self-avoiding random walks to 
construct $\phi^4_d$ models for $d \geq 2$, are briefly discussed in Section 
\ref{dimension}.

\subsection{$\phi^4_d$ models and their dependence on $d$} \label{dimension}

     According to the classification arising from the application of
standard perturbation theory to Lagrangian-based quantum field models,
the $\phi^4_d$ model is superrenormalizable when $d = 2,3$, renormalizable
when $d = 4$, and nonrenormalizable when $d \geq 5$. As seen above, the 
$\phi^4_2$ and $\phi^4_3$ models have been successfully constructed with many 
properties of physical relevance verified. However, the study of $\phi^4_4$ has 
made clear that there are further important subtleties to be understood in the 
construction of renormalizable quantum field models. Some of these are 
discussed in this section.

     It is a widespread view that quantum field models which are not 
asymptotically free (see below), such as $\phi^4_4$ with positive coupling 
constant, are not mathematically consistent. A close study of $\phi^4_4$ shows 
that this view must be nuanced. Another, closely related point is that 
perturbation theory is the primary tool used by field theorists to make 
predictions that may be checked against laboratory measurements. However, it 
is widely believed that perturbation series in QFT are {\it divergent}, at
least in models involving a bosonic field (this has been proven for 
$P(\phi)_2$ models by Jaffe \cite{Ja} and for the $\phi^4_3$ model by de Calan 
and Rivasseau \cite{CaRi}, but the initial suspicion that this should be true 
goes back to a simple heuristic 
argument of Dyson \cite{Dy} for QED). So one can hope that perturbation 
theory is {\it asymptotic} to an exact model, or even better, since the 
connection between the perturbation theory and the exact theory is then 
tighter as the quantities in the exact theory can be uniquely recovered by the 
Borel procedure from the corresponding perturbation series, the perturbation 
series is {\it Borel summable}. The Borel summability of the perturbation 
series for a number of quantities of physical interest has been verified in 
many of the 
models constructed to this date. However, an examination of $\phi^4_4$ is 
revealing also in this connection.\footnote{It should be emphasized that
even when the series is Borel summable, at some order of the expansion the
difference between the exact theory (if it exists) and the perturbation series 
prediction will get larger, not smaller, as one takes higher and higher orders
into account. The same is true of the difference between the experimentally
observed result and the perturbation series prediction.}

     Many authors have proven bounds which establish the local existence
of the Borel transform for the standard perturbation series for $\phi^4_4$
(\ie the Borel transform of the series exists and is analytic in a disk 
centered at 0), the most recent of which is due to Kopper \cite{Kop} (an 
overview and references to the earlier work may be found in \cite{Kop}, 
as well).

     Important insights have been won in QFT through the development of
{\it renormalization group} techniques. These have many concrete realizations,
some of which have been established in a mathematically rigorous manner.
We briefly describe one of the latter type in the specific case of $\phi^4_d$. 
Let $C_\kappa$ be a suitable ultraviolet cutoff free covariance such as
(\ref{cutoff}) or
$$C_\kappa = (-\Delta + m_0^2)^{-1} \; e^{-(-\Delta + m_0^2)/\kappa^2} \, ,$$
for which Euclidean square momenta larger than $\kappa^2$ are either strongly 
suppressed or totally eliminated and for which 
$C_\kappa \rightarrow (-\Delta + m_0^2)^{-1}$ in a suitable sense as 
$\kappa \rightarrow \infty$. The corresponding cutoff Euclidean measure is
$$\mu_{\lambda,\Lambda,\kappa} \equiv  \frac{e^{ -V(\lambda,\Lambda,\kappa)} \, \mu_{C_\kappa}}
{\int \, e^{ -V(\lambda,\Lambda,\kappa)} \, \mu_{C_\kappa}} \, , $$
where
\begin{equation} \label{renorm}
V(\lambda,\Lambda,\kappa) \equiv \int_\Lambda \, 
\lambda_\kappa Z_\kappa^2 :\phi_\kappa^4:(x) + 
\frac{1}{2} Z_\kappa \, \delta m_\kappa^2 :\phi_\kappa^2:(x) \; d^d x \, .
\end{equation}
This expression involves a wave function renormalization $Z_\kappa$, a mass 
renormalization $\delta m_\kappa^2$ and a coupling constant renormalization
$\delta \lambda_\kappa = \lambda_\kappa - \lambda$, all of which are
$\kappa$--dependent and some or all of which divergent as 
$\kappa \rightarrow \infty$, depending on the value of $d$. These quantities
can be given explicitly and are motivated by standard perturbation theory,
as above. As seen above, when $d = 2,3$ $Z_\kappa$ may be chosen to be 1
and the mass and coupling constant renormalizations are polynomials in
$\lambda$. When $d = 4$ the three renormalization counterterms are given
by power series in $\lambda$ which are likely to be divergent. They must
therefore be defined implicitly, and for this purpose renormalization
group techniques can be applied.

     The basic idea of the renormalization group applied in this context is to 
break up any integral with respect to $\mu_{\lambda,\Lambda,\kappa}$ into a sequence 
of integrals each over the field restricted to a particular range of momenta. 
To outline one concrete realization of this idea, let 
$\delta C = C_\kappa - C_{\kappa'}$ with $\kappa' < \kappa$ and 
$\delta \phi = \phi_\kappa - \phi_{\kappa'}$. This split of the field into
``high'' and ``low'' momentum parts $\phi_\kappa = \phi_{\kappa'} + \delta \phi$
leads to the replacement of $\mu_{\lambda,\Lambda,\kappa}$ with a suitable product
measure $\mu_{\lambda,\Lambda,\kappa'} \times \mu_{\lambda,\Lambda,\delta C}$. Carrying
out the integration over the measure on the right hand results in an
{\it effective action} $V_{\kappa'}(\phi_{\kappa'})$ determined by setting the
value of the said integration equal to 
$$e^{-V_{\kappa'}(\phi_{\kappa'})}\mu_{C_{\kappa'}}$$
up to a normalization factor. A computation determines the lower order in
$\lambda_\kappa$ contributions to $V_{\kappa'}$, and with a suitable choice of
$Z_{\kappa'}$ one finds that $V_{\kappa'}$ has, up to ``small'' terms, the 
same form as (\ref{renorm}).\footnote{These ``small terms'' are generally
ignored in heuristic QFT; however, showing that they can be controlled is
one of the main technical problems in the rigorous use of renormalization
group ideas.} The new renormalization terms 
$Z_{\kappa'},\delta m_{\kappa'}^2,\lambda_{\kappa'}$ can be computed as functions
of $Z_{\kappa},\delta m_{\kappa}^2,\lambda_{\kappa}$, resulting in the renormalization
group ``flow equations''.\footnote{In heuristic QFT the flow equations are 
usually expressed in terms of differential equations, \eg the Callan--Symanzik 
equation, which, however, have not been given a mathematically rigorous,
nonperturbative basis, in general.} In order that the ``small'' terms actually 
are small, it is necessary to choose $\kappa'$ suitably close to $\kappa$ and 
$\lambda_\kappa$ suitably small. Thus, beginning with a large value of $\kappa$
one must proceed in many incremental steps down from $\kappa_N = \kappa$ to a 
conveniently small $\kappa_0$ at which the final integration can be relatively
easily estimated to provide the desired bounds. But since the limit 
$\kappa \rightarrow \infty$ must ultimately be controlled, one would also need 
to have $\lambda_{\kappa_N} \rightarrow 0$ as 
$\kappa_N = \kappa \rightarrow \infty$, \ie as $N \rightarrow \infty$.
If this is so, then the model is said to be {\it asymptotically free} (in the
ultraviolet regime).

    In $\phi^4_4$ the flow equation for $\lambda_n \equiv \lambda_{2^n}$ is
$$\lambda_{n-1} \approx \lambda_n - \beta_2 \lambda_n^2 - \beta_3 \lambda_n^3$$
with $\beta_2 > 0$, so that for small $\lambda_n$ the preceding $\lambda_{n-1}$
is smaller than $\lambda_n$, contrary to the above picture. This is a signal
that the $\lambda \phi^4_4$ theory is not asymptotically free and the
renormalization group techniques cannot be applied in this case.

     However, the same flow equation shows that for $\lambda < 0$ this
procedure may be applicable. In fact, Gawedzki and Kupiainen \cite{GaKu2}
have carried out this procedure to provide a rigorous construction of
the (hierarchical) Euclidean $\lambda \phi^4_4$ model for negative coupling 
constant. The standard renormalized perturbation expansion is asymptotic
for the Schwinger functions. However, since reflection positivity is most 
likely not satisfied in this model, one does not arrive ultimately at a 
Minkowski space theory. Nonetheless, for spacetime dimensions less than 4 the 
corresponding flow equation for $\phi^4_d$ turns out to be consistent with the 
above picture also for {\it positive} coupling constant, and this approach 
yields another rigorous construction of the $\lambda \phi^4_3$ model for 
positive $\lambda$.

     Another approach to the study of $\phi^4_d$ models was motivated by 
Symanzik's insight \cite{Sy2} that Euclidean $\phi^4_d$ theory can be 
understood as a classical gas of weakly self-avoiding random paths and loops. 
To get some idea of this, consider the operator $-\Delta + m_0^2$ in its
approximation as a difference matrix on the lattice $\ZZ^d$, and write it
as a sum of diagonal and off-diagonal terms:
$$ \beta^{-1} I - J \equiv (2d + m_0^2)I - J \, , $$
where I is the identity matrix. The entries of the matrices $I,J$ are
indexed by the lattice sites in $\ZZ^d$, and the value of $J_{xy}$ is 1 if 
$xy$ is a lattice bond and 0 otherwise. So the covariance of the 
corresponding Gaussian lattice measure can be written as
$$(-\Delta + m_0^2)_{xy}^{-1} = (\beta^{-1} I - J)_{xy}^{-1} =
\sum_{n = 0}^\infty \, \beta^{n+1} (J^n)_{xy} \, .$$
The quantity $(J^n)_{xy}$ can be interpreted in this picture as a sum over
all possible walks along the lattice which go from site $y$ to site $x$
in $n$ steps. Observe that $\beta = (2d + m_0^2)^{-1} < 1$ and that $J_{xy}$
is exponentially damped as the distance between $x$ and $y$ grows to 
infinity. If $w$ is a path ( a ``walk'') along the lattice using nearest
neighbor bonds in the lattice (here viewed as a (hyper)cubic lattice) and 
$\vert w \vert$ is the number of lattice bonds in the walk $w$, then
the random walk representation of the covariance is
$$(-\Delta + m_0^2)_{xy}^{-1} = \sum_{w : y \rightarrow x} \; 
\beta^{\vert w \vert + 1} (J^{\vert w \vert})_{xy} \, ,$$
where the sum is over all walks from $y$ to $x$. Symanzik
showed how one could use this random walk representation for the two-point
Schwinger function of the free field to give an expression for the Schwinger
functions of the interacting model (on the lattice, in finite volume).
This results in studying correlations of random paths, where the weight
functions on the random paths is significantly more complicated than the
straightforward exponential weight above \cite{Sy2}. A multitude of different
random walk representations is to be found in the literature - see the
monograph by Fernandez, Fr\"ohlich and Sokal \cite{FeFrSo} for a unified
presentation of many of these.

     Typically, the random walk formalism is used in a finite volume, lattice 
approximation of the model to derive various kinds of correlation inequalities, 
which then generally carry over easily to the infinite volume limit. These 
correlation inequalities are utilized to establish bounds on the continuum 
limit, \ie the exact model, and on the critical exponents of the model. 
Using these techniques, it has been definitively established by Aizenman 
\cite{Ai} and Fr\"ohlich \cite{Froh4} that for $d > 4$ the exact limit theory 
is trivial, \eg the exact model is Gaussian, manifest no interaction, 
has trivial scattering. On the other hand, in the case 
of $d = 4$ the situation is more subtle, as the limit may depend on the way 
the limits are taken and which details enter into the particular mode of 
construction --- the reader is referred to \cite{FeFrSo} for details. In many 
of these circumstances it has been proven that the exact theory is again 
trivial in four spacetime dimensions. This is striking, since the standard
renormalized perturbation series for the model exists to all orders and
is not trivial. Hence, the perturbation series is {\it not} asymptotic to
what is apparently the exact theory. Moreover, the classical limit of that 
exact quantum theory could not coincide with the classical
$\phi^4_4$ field theory.

     This brings us to the question: which criteria do we use to decide
when a given mathematical model is a/the M$_d$ quantum field theory?
For most of the models discussed to this point one has the reassurance that
the standard (possibly renormalized) perturbation series is asymptotic
to some set of important quantities, \eg the Schwinger functions, in the
``exact model'' (and in many cases the latter quantities can be uniquely
recovered from the Borel transform of the former series). For some of these 
theories one has even been able to prove that in the ``exact model'' the 
fields satisfy the (suitably interpreted) semiclassically motivated field 
equations. For further reassurance, some researchers would also like to know
that (A) the classical limit of the ``exact model'' coincides with (B) the 
classical theory associated with the Lagrangian selected by whatever 
semiclassical reasoning went into the choice of the model. This latter
question has not yet received much attention from mathematical physicists,
but Donald \cite{Do} has established a result for $P(\phi)_2$ models
with convex polynomial $P$ which relates the classical limit of certain
quantities in the quantum field model to corresponding quantities in the
classical model. However, the relation between (A) and (B) is not
a 1-1 correspondence, even in the well-behaved $P(\phi)_2$ models. For
example, Slade \cite{Sl} has shown that for values $a$ of the classical
field for which the classical potential $U_0(a) = P(a) + \frac{1}{2} m^2 a^2$
does not equal its convex hull, the correspondence between (A) and (B)
can break down.

     As already seen, none of these desiderata is satisfied by the (trivial) 
``exact model'' for $\phi^4_4$. These and other considerations have motivated 
Klauder to propose an alternative way to construct ``exact'' $\phi^4_4$. The 
basic idea is not to perturb the free measure, introduce counterterms 
motivated by standard perturbation theory and then control a number of limits, 
as has been done in the work indicated above, but to perturb a ``pseudofree'' 
measure, introduce a different set of counterterms and then control certain 
limits to get the ``exact model'' \cite{Kl2}. This program has not been 
completed, so the reader is referred to \cite{Kl1,Kl2} for more details.  

\subsection{Yukawa$_d$ models} \label{yukawa}

    Theories with fermions are amenable to Euclidean methods, as well, if
the interactions are quadratic in the fermions \cite{OS0}, including
therefore Yukawa-type interactions, \ie interactions of the form
$\overline{\psi} \Gamma \psi \phi$, 
$\Gamma = 1, i \gamma_5 = - \gamma_0 \gamma_1$, for the scalar, 
respectively pseudoscalar, coupling . Initially, the Fermi fields were
``integrated out'', following Matthews and Salam, resulting in an effective 
action involving only the boson field called the Matthews-Salam 
determinant. Due to the necessity for a mass renormalization, the renormalized 
Matthews--Salam determinant is given by
$$\det{}_{\text{ren}}(1 + \lambda K(\phi)) \equiv
\det{}_3( 1 + \lambda K(\phi)) 
\exp\left[ \frac{\lambda^3}{3} \text{tr} K^3(\phi) +
\delta m^2 \int :\phi^2:(x) \, d^2x \right] \, , $$
where 
$$\det{}_n(1 + A) \equiv \det\big[ (1+A)e^{\sum_{k=1}^n \frac{1}{k} (-A)^k} \big] \, ,$$
$K = S \Gamma \phi$, $S$ is the Euclidean fermion
propagator and $\delta m^2$ is a boson mass counterterm
arising from second order perturbation theory. The ultraviolet cutoff index
$\kappa$ has been suppressed. Seiler \cite{Se} showed that, after introducing
a suitable finite volume cutoff into $K$ and $\int :\phi^2:(x) \, d^2x$, 
the renormalized determinant has
the necessary integrability properties with respect to the free boson measure
$\mu_C$ as $\kappa \rightarrow \infty$. Subsequently, Magnen and 
S\'en\'eor \cite{MaSe2} and Cooper and Rosen \cite{CoRo} independently 
verified the Osterwalder--Schrader axioms for the Schwinger functions of the 
infinite volume limit by using cluster expansions. The Fermi fields in the
Schwinger functions are also ``integrated out'':
the Schwinger function for $n$ boson fields, $m$ fermion and $m$ anti-fermion
fields is given by
\begin{equation} \label{MS}
Z^{-1} \int \big( \prod_{l = 1}^n \phi(x_l) \big) \; 
\det \big[ S^\prime (y_i,z_k;\phi) \big] \, 
\det{}_{\text{ren}}(1 + \lambda K(\phi)) \; d\mu_C(\phi) \, ,
\end{equation} 
where 
$$Z \equiv \int \det{}_{\text{ren}}(1 + \lambda K(\phi)) \; d\mu_C(\phi) \, ,$$ 
the determinant is applied to the matrix whose $(i,j)$th entry is
$S^\prime (y_i,z_k;\phi)$, $i,j = 1,\dots,m$, which is the two point Schwinger 
function for the fermions in the external field $\phi$ and determined by 
$(1 + \lambda K) S^\prime = S$ (all cutoffs are suppressed for transparency).

     Renouard \cite{Re} proved that the perturbation expansion for 
the Schwinger functions of the limit theory is Borel summable. Balaban and 
Gawedzki \cite{BaGa} established the existence of a phase 
transition in the two-dimensional Euclidean pseudoscalar Yukawa model for 
sufficiently large
fermion mass. To do so, they adapted the mean field cluster expansion 
method developed by Glimm, Jaffe and Spencer \cite{GJS1} to this
situation.

     Lesniewski \cite{Le} utilized advances in CQFT developed for purely
fermionic models to reverse the initial procedure. He ``integrated out''
the bosonic field first, resulting in an effective fermionic action of the 
form
$$ \frac{1}{2} \int \, g_\kappa(x-y) \, :\overline{\psi} \psi:(x) \, 
:\overline{\psi} \psi:(y) \, dx dy \, , $$
where
$$g_\kappa(x-y) \equiv \frac{\lambda^2}{(2\pi)^2} 
\int \frac{e^{ip(x-y)}}{p^2 + m^2 + \delta m_\kappa^2(\lambda)} \, .$$
$\lambda$ is the coupling constant and the mass counterterm  
$\delta m_\kappa^2(\lambda)$ is given by second order perturbation theory.
Using renormalization group techniques developed to treat the Gross-Neveu
model (see the next section), he reproved the existence and the Borel 
summability of the limit theory.

     By modifying the phase space cell expansion developed by Glimm and 
Jaffe to deal with the $\phi^4_3$ model, Magnen and S\'en\'eor \cite{MaSe4}
established the Osterwalder--Schrader axioms and the Borel summability of the
standard perturbation theory for the Schwinger functions of the pseudoscalar
Yukawa$_3$ model and also indicated how to prove the corresponding results for
the scalar Yukawa$_3$ model.

\subsection{Gross--Neveu$_2$ model} \label{GN}

     The Lagrangian for the (massive) Gross--Neveu model is given by
$$ \overline{\psi}(x) \, ( i\zeta \slashed\partial + m) \psi(x) +
\frac{\lambda}{N} \big( \overline{\psi}(x) \psi(x) \big)^2  \, ,$$
where $\psi$ is a fermion field with $N$ components (colors) and $m > 0$ is
the fermion mass. In the original model proposed by Gross and Neveu in
1974 \cite{GrNe}, the bare mass was 0, but the field ``acquired a mass'' by a 
complicated mechanism called {\it dynamical symmetry breaking}. This heuristic 
picture has been partially supported by results by Kopper, Magnen and Rivasseau 
for sufficiently large $N$ \cite{KoMaRi}, but we shall restrict the discussion
to the massive case, for which definitive results have been attained. 

     In two spacetime dimensions the model is renormalizable and asymptotically 
free when $N > 1$. The (somewhat simplified) spatially and ultraviolet cutoff 
action is given by
$$\int_\Lambda \, \big[ 
\frac{\lambda}{N} \big( \sum_{a=1}^N \, \overline{\psi}_a(x)\psi_a(x) \big)^2  + 
\delta m \, \big( \sum_{a=1}^N \, \overline{\psi}_a(x)\psi_a(x) \big) +
\delta \zeta \, \big( \sum_{a=1}^N \, \overline{\psi}_a(x) \, i\slashed\partial 
\psi_a(x) \big) \big] \; d^2x $$
where $\delta m$ is the mass counterterm, $\delta \zeta$ the wave function
counterterm, and $N > 1$. The spinor indices and the ultraviolet cutoff 
subscripts are suppressed for transparency. 

     Taking advantage of the fact that perturbation theory for fermions is
much simpler than that for bosons, Gawedzki and Kupiainen \cite{GaKu1},
on one hand, and Feldman {\it et alia} \cite{FMRS}, on the other, provided 
different proofs that the Schwinger functions converge, as the volume and 
ultraviolet cutoffs are removed, to Schwinger functions satisfying the 
Osterwalder--Schrader axioms for all sufficiently small values of the 
renormalized coupling constant. Moreover, the limit Schwinger functions are
the Borel sum of their standard renormalized perturbation series \cite{FMRS},
\ie the expansion in the renormalized coupling constant. By employing more 
recent advances in the rigorous treatment of renormalization
group methods, Disertori and Rivasseau \cite{DiRi} have found a significantly
simpler proof of these results and have shown that the mentioned Borel
summability is uniform in $N$. Moreover, they demonstrate that the 
renormalization group equations and the $\beta$ function are well defined
in the fully interacting limit theory. Iagolnitzer and Magnen \cite{IaMa} 
studied the Bethe--Salpeter kernel in this model and proved two-particle
asymptotic completeness in this model. 

     The Gross--Neveu model has also been studied in 3 spacetime dimensions,
where it is neither renormalizable nor asymptotically free. Heuristically,
however, it is asymptotically free in the limit $N \rightarrow \infty$.
After some changes of variables to represent the model for large $N$ as a 
perturbation of its $N \rightarrow \infty$ limit and a rigorous summation
of the simplest, most divergent coupling constant contributions that changes 
the nonrenormalizable model into a renormalizable one, de Calan, Faria de Veiga,
Magnen and S\'en\'eor have shown that for sufficiently large $N$ the Schwinger 
functions of GN$_3$ exist after the ultraviolet and volume
cutoffs are removed, though most of the Osterwalder--Schrader axioms have
not yet been verified (\cf \cite{Ca}). It is 
striking that the exact Schwinger functions exist but the standard
perturbation series in this model does not, since it is not renormalizable.

\subsection{Sine--Gordon$_2$ and Thirring models} \label{sG} 

     In this section we discuss a set of models related to the 
sine--Gordon model, which is a model of a real, scalar massive or massless 
bosonic field $\phi$ in two spacetime dimensions with (spatially cutoff) 
action
$$V_\Lambda \equiv \zeta \int_\Lambda \; : \cos (\alpha\phi + \theta):(x) 
\; d^2x \, ,$$
where $\alpha, \zeta \in \RR$ and $\theta \in [0,2\pi)$. A number of different 
parametrizations of this action are to be found in the literature.

     The ultraviolet divergences of this model depend on the size of
$\vert \alpha \vert$. If $\alpha \in (-2\sqrt{\pi},2\sqrt{\pi})$, the model
is superrenormalizable and already the measure
$$\mu_\Lambda \equiv \frac{e^{V_\Lambda} \mu_C}{\int e^{V_\Lambda(\phi)} \, d\mu_C(\phi)} 
$$
is well defined. In this range for $\alpha$ the model is amenable to the 
techniques previously developed to construct $P(\phi)_2$ models and models 
with exponential interaction in two spacetime dimensions. Fr\"ohlich and
Seiler \cite{FrSe} proved that for sufficiently small 
$\vert \zeta \vert /m_0^2$ the standard perturbation series expansion in $\zeta$
for the Schwinger functions actually converges in the infinite volume limit,
the Schwinger functions satisfy the Osterwalder--Schrader axioms and the mass 
hyperboloid is isolated. The Haag--Ruelle scattering theory is applicable,
and the S--matrix is nontrivial. Moreover, the perturbation series in $\zeta$
is asymptotic to the scattering amplitudes.

     On the other hand, Park \cite{Pa2} showed that for the massless 
sine--Gordon model the infinite volume limits of the expectations (with respect
to the corresponding finite volume measures) of
products of (smeared) fields of the form $: \cos \epsilon(\phi + \theta):(x)$,
$: \sin \epsilon(\phi + \theta):(x)$ and ${\bf a} \cdot {\bf \nabla}\phi(x)$
(${\bf a}$ a constant vector) satisfy the Osterwalder--Schrader axioms 
(with the possible exception of clustering) for all parameter values in the 
indicated ranges by using suitable correlation inequalities. 

     When $\alpha^2 \in [4\pi,8\pi)$ the model is superrenormalizable, but the
number of renormalization counterterms increases to infinity as $\alpha$ 
increases to $8\pi^2$. Nicol\`o, Renn and Steinmann \cite{NiReSt} proved that 
in a finite volume the massive sine--Gordon model is ultraviolet stable for 
$\alpha$ in this range. In the same range Dimock and Hurd use renormalization 
group methods to prove bounds on the finite volume Schwinger functions 
for both the massless and massive cases which are uniform in
$\Lambda$ in the massive model. Moreover, they show that in a finite volume
the Schwinger functions are analytic in $\zeta$ at 0. However, the full
control of the infinite volume limit has not yet been attained.  

     When $\alpha^2 = 8\pi$ the model is heuristically renormalizable but not 
superrenormalizable, and when $\alpha^2 > 8 \pi$ it is nonrenormalizable.
Nicol\`o and Perfetti \cite{NiPe} have proven that for $\alpha^2 = 8\pi$
the model is indeed perturbatively renormalizable. And for $\alpha^2 > 8 \pi$
Dimock and Hurd \cite{DiHu} have shown that the model is asymptotically
free in the infrared. However other rigorous results in these cases are not 
yet available.

     One of the first manifestations of the equivalence between apparently
distinct quantum field models is the phenomenon of {\it bosonization} in
two spacetime dimensions. A simple example is the equivalence between
free massless fermionic fields and free massless bosonic fields given
by the identifications
$$\overline{\psi}(1 + \sigma \gamma_5)\psi \leftrightarrow c \, 
: e^{i\sigma \sqrt{4\pi} \phi} : \; , \; 
\overline{\psi} \gamma^\mu \psi \leftrightarrow 
-\frac{1}{\sqrt{\pi}}\epsilon^{\mu\nu} \partial_\nu \phi \, ,$$
where $\sigma = \pm 1$, $\epsilon$ is the standard totally antisymmetric
matrix with entries $\pm 1, 0$, and $c$ is a suitable constant depending
on the choice of Wick product used. Coleman \cite{Co} gave a heuristic
argument suggesting a similar equivalence between the massive Thirring
model with Lagrangian
$$Z \overline{\psi} \, i\slashed\partial \psi - 
\tau Z_1 \overline{\psi} \psi - \frac{\lambda}{4} Z^2 j_\mu j^\mu \, ,$$
where $Z,Z_1$ are renormalization constants and 
$j_\mu = \overline{\psi} \gamma^\mu \psi$, and the massless Sine-Gordon
model with Lagrangian
$$ \frac{1}{2} \partial_\mu \phi \, \partial^\mu \phi + 
\zeta : \cos(\alpha \phi): $$
under the identifications
$$Z_1 \overline{\psi}(1 + \sigma \gamma_5)\psi \leftrightarrow c \, 
: e^{i\sigma \alpha \phi} : \; , \;
Z \overline{\psi} \gamma^\mu \psi \leftrightarrow 
-c_1 \epsilon^{\mu\nu} \partial_\nu \phi \, ,$$
where $c,c_1$ are suitable constants depending on $\lambda$ and the ultraviolet
cutoff used. In order for this equivalence to be valid, certain relations
among the Thirring parameters $\lambda,\tau$ and the Sine--Gordon parameters
$\zeta,\alpha$ must obtain. The case $\alpha^2 = 4\pi$ corresponds to free
fermions ($\lambda = 0$), and the case $\zeta = 0$ (free bosons) corresponds
to massless fermions ($\tau = 0$).   

     Closely associated with these models is the Thirring--Schwinger model,
often referred to as QED$_2$ since the only manifestation of electromagnetism
in one spatial dimension is the Coulomb force between charges. The 
interaction Lagrangian for this model is
$$ \frac{g}{2} j_\mu j^\mu + 
\frac{\pi e^2}{2} \widetilde{j}^0 V \widetilde{j}^0 \, , $$
where $\psi$ is a massive two-component Fermi field, $j_\mu$ is the 
conserved current defined above, and $\widetilde{j}^0 = j^0 + j_c^0$,
with $j_c^0$ is a formal $c$-number current specifying charges at infinity,
and $V = \frac{1}{2} \vert x \vert$ is the one-dimensional Coulomb potential. 
Fr\"ohlich and Seiler \cite{FrSe} rigorously proved the equivalence between
this model and the massive sine-Gordon model for sufficiently large mass
and $\alpha^2 < 4\pi$ by showing, along the lines of Coleman's original
argument, that their perturbation series coincide term by term and actually
converge.
  
     From the point of view of heuristic Lagrangian considerations, the 
Thirring model is renormalizable but not superrenormalizable. In point of fact, 
different versions of the (massless) ``Thirring model'' are extant. There is 
the original model introduced by Thirring \cite{Th}, for which Glaser 
\cite{Gl} found an explicit ``solution'' for the fields. However, 
Ruijsenaars \cite{Ru} showed that the $n$-point 
functions of the fields of this ``solution'' do not exist. Another concrete 
realization was proposed by Johnson \cite{Joh}, which, although it is
mathematically unsatisfactory, led to a rigorous formula by Klaiber \cite{Kla}
for Wightman functions which coincide with the $n$-point functions of
Johnson's realization for $n = 2,4$. Carey, Ruijsenaars and Wright
\cite{CaRuWr} proved that Klaiber's functions satisfy the Wightman axioms.
In fact, they rigorously constructed the fields corresponding to these 
$n$-point functions as strong limits of certain natural approximating fields.

     A rigorous construction of the massive Thirring model was finally
achieved by Benfatto, Falco and Mastropietro \cite{BeFaMa1}. Writing the
ultraviolet cutoff generating functional for the Euclidean model as a 
Grassmanian integral, they showed that after proper choice of the wave function
renormalization and bare mass the Schwinger functions (at noncoinciding
points) converge as the cutoff is removed to Schwinger functions satisfying
the Osterwalder--Schrader axioms. A multiscale renormalization group approach
related to the one discussed in Section \ref{dimension} is used in the proof,
and so the results are valid for sufficiently small values of the coupling
constant (there is no restriction on the mass). Curiously, they
also find that in the massless case the resultant two-point function differs
from that found by Johnson. Benfatto, Falco and Mastropietro \cite{BeFaMa2}
then went on to prove Coleman's equivalence between the massless sine--Gordon 
model and the massive Thirring model in any finite volume. 

\subsection{Local gauge quantum field theories}  \label{gauge}

     Local gauge quantum field theories\footnote{The adjective ``local'' 
modifies ``gauge'', not ``quantum field theory''. The internal symmetry groups
in these theories are infinite dimensional, as the ``gauge transformation''
can depend upon the spacetime point at which it is being implemented. This is
to be contrasted with ``global gauge groups'', such as the global U(1) symmetry
in the Yukawa models or the global $\ZZ_2$ symmetry in $P(\phi)_2$ models
with even polynomial $P$.} are conceptually and mathematically quite 
challenging, and we discuss briefly some of these challenges in this section. 
Although we are concerned here with CQFT, it is necessary to treat some of the 
insights gained into the nature of gauge theories by other mathematically 
rigorous means, since they shed light on the nature of these challenges and
on the consequences for the mathematical framework of the models. But even the 
mathematically rigorous literature on aspects of local gauge theory has become 
enormous, so only a few highlights can be discussed here.

     One of the challenges facing the rigorous construction of such models
is to decide what constitutes the proper framework for the end result.
It has become clear that, though the Wightman axioms are still 
suitable for the {\it gauge invariant local} fields in gauge theories, such 
as the electromagnetic fields $F^{\mu\nu}$, they must be supplemented to 
include extended field objects, since it is also useful to consider gauge 
invariant objects such as 
\begin{equation} \label{loop}
\overline{\psi}(x) \; e^{i \int_{C_{xy}} \, A_\mu(z) \, dz^\mu} \; \psi(y) \, ,
\end{equation}
here $C_{xy}$ is a suitable curve connecting the points $x$ and $y$, $A$
is a gauge potential, and $\psi$ a fermionic field. Various incomplete 
proposals have been made in this regard, but the most developed of these 
appears to be that of Fr\"ohlich, Osterwalder and Seiler (\cf the last chapter 
of \cite{Se2}), which provides conditions on expectations of products of 
Euclidean Wilson loops $e^{i \int_C \, A_\mu(x) \, dx^\mu}$ for piecewise smooth loops 
$C$ in spacelike hyperplanes in Minkowski space (viewed, however, in Euclidean 
space) as well as a procedure to construct the corresponding Minkowski space 
theory. Another proposal by Ashtekar, Thiemann and co-workers \cite{ALMMT,Thi} 
places its conditions instead on the Euclidean measure of the model, instead
of on specific classes of expectations. These conditions also afford 
the possibility to reconstruct important aspects of the real time theory
from the Euclidean data. They have shown that a large class of Yang--Mills
models in two spacetime dimensions verify these conditions --- see Section
\ref{YM}.

     In addition, it is convenient for various purposes to include the 
unobservable gauge potentials $A_\mu$ in heuristic gauge QFT. But if they are 
allowed to enter into the theory as a quantum field in their own right, the 
price is high. It was recognized by many physicists that the introduction of 
gauge potentials as quantum fields acting on the state space is incompatible 
with both Lorentz covariance and Einstein causality. We illustrate this point
with the electromagnetic field in Section \ref{qed}. The upshot is that
either one must choose a gauge in which the gauge potential acts as an 
operator in a Hilbert space but is not Lorentz covariant and violates Einstein 
causality, or one must choose a gauge in which the gauge potential is 
Lorentz covariant and satisfies Einstein causality but acts as an operator 
in a vector space with an indefinite inner product, \ie there exist vectors 
$\Psi$ in the state space such that $\langle \Psi, \Psi \rangle < 0$. In such 
a circumstance the standard relation between quantum expectations and 
probabilities fails. One should note that perturbation theory is generally
performed in gauges of the latter type. Additional complications are
introduced by the (again unobservable) charge carrying Fermi fields in such 
quantum field theories (again illustrated in Section \ref{qed}). 

     A modified version of the Wightman axioms which takes into account the
necessity of employing a reasonably behaved indefinite inner product 
space called a Krein space may be found in the monograph of Strocchi 
\cite{Str}.\footnote{A modified version of the Osterwalder--Schrader axioms
for the same purpose has been proposed by Jakobczyk and Strocchi \cite{JaStr}.
They show that if the Schwinger functions satisfy these modified conditions,
then the reconstructed Wightman functions satisfy the modified Wightman 
axioms.} The essential modification is that the positivity condition on the 
Wightman functions (or the reflection positivity condition on the Schwinger
functions) is replaced by a ``Hilbert space structure'' condition 
which permits the construction of subspaces of the Krein space
suitably associated with the Wightman functions on which the inner product is
positive definite.

     Taking the standpoint that the Euclidean gauge potentials are to be taken
explicitly into account as dynamical variables brings up the question of
which measure to adopt on the space of potentials. In the geometric formulation
of classical electrodynamics, the gauge potentials are connections on a
certain fiber bundle with base space $\RR^d$ (and the Wilson loops are (traces 
of) holonomies of the connection around closed loops). Let $\mathscr{A}$ denote 
the set of such connections, and let $\mathscr{G}$ denote the local gauge 
group acting on $\mathscr{A}$. In gauge theories it is natural to view each 
gauge equivalence class of connections as a distinct physical ``path'', \ie 
the orbit of any single potential $A \in \mathscr{A}$ under the action of 
$\mathscr{G}$ is an equivalence class. The elements of each orbit are thus 
viewed as physically equivalent. However, the resultant path space --- the 
space of all such orbits --- is a nonlinear quotient space 
$\mathscr{A}/\mathscr{G}$; this introduces technical difficulties. By using a 
gauge fixing one can impose a linear structure on 
$\mathscr{A}/\mathscr{G}$ for $d = 2$, but in higher dimensions, the 
resultant Gribov ambiguities limit the usefulness of such gauge fixings. 

     A further complication arising when one takes the gauge potentials as
dynamical variables is that the measures of physical relevance in QFT are 
typically not supported on functions having nice properties from the point of 
view of analysis (in particular, the measures are not supported on 
$\mathscr{A}/\mathscr{G}$), so that one must introduce a suitable
closure of $\mathscr{A}/\mathscr{G}$ in which one can find the relevant
generalized connections on which these measures do have support. The choice of 
this closure is not unique, and many can be found in the literature. The 
measures are then defined on this closure. In the scalar and fermionic 
Euclidean models described above, the (cutoff) physical measure is defined as 
a suitable perturbation of a Gaussian measure corresponding to a free field.
In CQFT this Gaussian measure replaces the ``measure'' 
$\exp \{ -m^2_0 \int \phi^2(x) \, dx^d \} \, d\nu(\phi)/Z$ to be found in 
heuristic treatments of functional integration in QFT, where $\nu$ is the 
nonexistent Lebesgue measure on $\Ss^\prime (\RR^d)$. 
For gauge models Ashtekar and Lewandowski \cite{AsLe} have 
constructed a uniform Borel measure $\mu_0$ on a natural closure of 
$\mathscr{A}/\mathscr{G}$ which is gauge invariant and 
strictly positive on continuous cylindrical functions. This measure replaces
$\nu$ in many rigorous treatments of functional integration for gauge theories.
The physical measures are then obtained by suitably perturbing $\mu_0$. This 
is illustrated in Section \ref{YM}. The advantage of this approach is that the 
gauge invariance is explicit at every step. Any Fermi fields coupled to the 
gauge potentials would be ``integrated out'' along the lines of the previous 
work on the Yukawa models, resulting in a new effective action perturbing the 
pure gauge measure.

     The complications in defining measures on such path spaces are great,
leading many researchers to concentrate on the topological and
geometric aspects of the problems (in so-called topological quantum field 
theory), largely bypassing (or ignoring) the analytical aspects. These 
interesting developments are not treated here. And, although also classical
gravitation can be understood as a local gauge theory, we do not address
any aspect of quantum gravitation here. However, we give a brief accounting of 
the constructive results concerning the physically most important quantum gauge 
models, even though definitive results have been attained only in 
spacetime dimensions less than 4 and, as indicated above, aspects of these 
models do not conform to the Wightman setting. It should be emphasized that it 
is not at all clear that gauge theories such as QED and QCD cannot be 
incorporated into the HAK setting, since there the observable quantities are 
primary and many of the problems we have discussed above arise only after 
introduction of unobservable fields. One of the many successes of AQFT is that 
Doplicher, Haag and Roberts have shown how, starting with the (necessarily 
gauge invariant) observables in a {\it global} gauge theory, one can uniquely 
{\it derive} both the gauge group and the associated charge carrying fields 
--- \cf \eg \cite{Ar,Haag}. Although a corresponding breakthrough for 
{\it local} gauge theories has not yet been achieved, the possibility remains
open.

\subsubsection{Abelian Higgs$_d$ model} \label{higgs}

     The abelian (or $U(1)$) Higgs model is an interacting theory of a 
vector field $A_\nu(x)$ coupled in a gauge covariant manner to an 
$N$-component scalar field $\phi(x)$ and is sometimes referred to as
scalar electrodynamics (when $N = 1$). The Euclidean action of the model in
$d$ spacetime dimensions is
$$\int \; \big( \frac{1}{4} \sum_{\mu,\nu = 1}^d \vert F_{\mu\nu}(x) \vert^2 +
\frac{1}{2}\sum_{\mu = 1}^d \vert D_\mu \phi(x) \vert^2 +
\frac{1}{2} m^2 \vert \phi(x) \vert^2 + \lambda \vert \phi(x) \vert^4 \big)
\; d^d x \, ,$$
where the field strength tensor is 
$F_{\mu\nu}(x) = \partial_\mu A_\nu - \partial_\nu A_\mu$, and the covariant 
derivative of the scalar field is
$$(D_\mu \phi)_i(x) = \partial_\mu \phi_i(x) - e A_\mu(x) (Q\phi)_i(x) \, . $$
$Q$ is an antisymmetric $N \times N$ matrix, and $e$ and $\lambda$ are coupling
constants. For $d = 2,3$ the model is superrenormalizable.

     In a series of papers (see \cite{BrFrSe} for references), Brydges,
Fr\"ohlich and Seiler studied this model in two spacetime dimensions by 
commencing with the model on a Euclidean lattice. For the convenience
of avoiding spurious infrared divergences, the bare mass of the gauge field 
$A_\mu$ is initially taken to be strictly positive, but ultimately the 
continuum limit, the ultraviolet limit and the limit in which the bare mass of 
the gauge field tends to zero are all controlled, and all the 
Osterwalder--Schrader axioms except clustering are verified for Schwinger
functions involving {\it gauge invariant} local fields such as 
$: \vert \phi \vert^2$ and $F_{\mu\nu}$, as well as string and loop observables
such as 
$$:\overline{\phi}(x) \; e^{\int_x^y \; A_\mu \, dx_\mu^\prime} \; \phi(y): $$
and $:e^{\oint \; A_\mu \, dx_\mu}:$. Hence, there exists a corresponding Wightman
theory for these fields, though the vacuum may not be unique.

     King \cite{Ki1,Ki2} studied the abelian Higgs model in two and three
spacetime dimensions. He also began with the model on a finite volume lattice
and controlled the continuum and infinite volume limits using a combination
of renormalization group techniques and correlation inequalities. He verified
all of Osterwalder--Schrader axioms except ergodicity and the regularity 
properties which assure that the Wightman functions are tempered distributions.

     Balaban, Imbrie and Jaffe \cite{BaImJa} have examined the abelian
Higgs model for $d =2,3$ for rigorous evidence supporting the heuristic
notion of the Higgs mechanism, \ie the notion that, under certain circumstances
usually associated with spontaneous breaking of the gauge symmetry, a massless 
particle can acquire mass. A proof of the existence of a mass gap 
on the Hilbert space generated from the vacuum by application of all gauge 
invariant observables could be taken as such supporting evidence. Together
with Brydges they supplied such a proof for the model on a lattice. Although
they made much progress on the proof of the mass gap in the continuum limit
\cite{BaImJa}, the final step in that proof never appeared in print.

\subsubsection{Quantum electrodynamics$_d$} \label{qed}

     Despite the success of the experimental predictions made by the
perturbation theory computations associated with QED, it is widely believed
that since QED is not asymptotically free in four spacetime dimensions 
it cannot be defined as a mathematically rigorous theory; but this matter 
has not been settled. Moreover, beyond the ultraviolet problems inherent 
in the model some serious conceptual difficulties remain to be resolved. As a 
first example, though the notion of a gauge transformation is straightforward 
in classical electrodynamics, it is not at all clear what gauge transformations 
are in QED\footnote{as evidenced, for instance, by an investigation by 
Strocchi and Wightman \cite{StrWi}} and what their relation may be to the 
classical gauge transformations which play such an important role in the 
semi-classical reasoning commonly found in heuristic QFT. 

     In addition, as shown by a number of mathematical physicists (see 
Strocchi's monograph \cite{Str} for details and references), under various 
sets of reasonable assumptions, the standard picture in quantum theory of a 
Hilbert space $\Hs$ serving as the state space for the model is inconsistent 
with each of the following: the potential field $A_\mu$ is covariant, $A_\mu$ 
satisfies Einstein causality, and Maxwell's equations are satisfied on $\Hs$. 
Wightman and G\aa rding \cite{WiGa} have shown that for the free 
electromagnetic field a formalism due to Gupta and Bleuler is mathematically 
consistent. In particular, there exists a vector space $\Hs$ with a 
sesquilinear Hermitian form $\langle \cdot, \cdot \rangle$ on which acts a 
unitary representation $U(\Pid)$ of the Poincar\'e group satisfying the 
spectrum condition and operator valued tempered distributions $A_\mu(x)$ and
$F_{\mu\nu}(x) = \partial_\mu A_\nu(x) - \partial_\nu A_\mu(x)$, which are covariant
under the action of $U(\Pid)$ and satisfy Einstein causality. In $\Hs$ is
a distinguished subspace $\Hs'$ such that $\langle \Psi,\Psi \rangle \geq 0$
for all $\Psi \in \Hs'$ and in which Maxwell's equations hold:
$$\langle \Phi, \partial^\mu F_{\mu\nu}(x) \Psi \rangle = 0 \, $$
for all $\Phi,\Psi \in \Hs'$. The (pure) physical states are described
by unit vectors in the Hilbert space given by the quotient $\Hs'/\Hs''$,
where $\Hs''$ is the subspace of $\Hs'$ consisting of vectors $\Psi$
such that $\langle \Psi,\Psi \rangle = 0$. For these the standard
quantum probability interpretation is valid. 

     Though there are a number of realizations of the Euclidean free 
electromagnetic field in spacetime dimension $d$, the most concise is to 
specify its generating function:
\begin{equation} \label{min}
f \mapsto \exp \big[ -\frac{g^2}{2} 
\langle (-\Delta_d)^{-1} \delta f, \delta f \rangle \big] \, ,
\end{equation}
where $\Delta_d$ is the Laplacian on $\RR^d$, 
$(\delta f)_\nu(x) = \sum_\mu \partial f_{\mu\nu}(x)/\partial x^\mu$ and
$f$ is a tempered function valued 2-form. This satisfies the hypotheses
of Minlos' theorem, so there exists a probability measure $\mu$ on the dual
space of the space of test 2-forms such that $\int \, \exp [iF(f)] \, d\mu$
equals (\ref{min}). The Euclidean free electromagnetic field can be defined
as the generalized stochastic process $f \mapsto F(f)$ satisfying that
equation. The corresponding Schwinger functions satisfy all of the
Osterwalder--Schrader axioms (note that the gauge potentials do not appear
at all). The exterior derivative of $F$ is zero. Realizations of the
Euclidean free electromagnetic field involving the gauge potentials run into 
the problems indicated above.

     Additional complications are introduced when one considers the electron
field $\psi(x)$. Because the photon is massless and Gauss' Law holds,
the physical electron field carrying the electric charge cannot be Lorentz
covariant or satisfy Einstein causality. Moreover, in a Hilbert space $\Hs$ the 
electron field $\psi(x)$ cannot satisfy Einstein causality if Maxwell's 
equations with a source hold in $\Hs$. Furthermore, in fully interacting QED
the physical state space will be complicated not only by the considerations
mentioned above and by the superselection rule associated with the global 
electric charge operator, resulting in countably infinitely many distinct 
charge sectors, but also by the uncountably infinitely many sectors associated 
with distinct asymptotic configurations of soft photons. (For a profound
investigation of these matters, the reader is referred to a paper of 
Buchholz \cite{Bu0}.) In heuristic QED this abundance of superselection 
sectors is ignored, and one selects more or less arbitrarily a subset of 
physical states to work with, typically discarding in this manner the charged 
states with the best possible localization properties. 

     The perturbation theory for QED is on solid mathematical footing. QED is 
renormalizable in four spacetime dimensions (see, in particular, the 
proof in \cite{FHRW}), so the perturbation series is well defined to all
orders. Indeed, it has been proven by Feldman and co-authors \cite{FHRW}  
that the standard perturbation series for the Schwinger functions in QED$_4$ 
is locally Borel summable. It is not known if there exists an exact
model to whose Schwinger functions this series is asymptotic. So we turn
now to lower dimensions.

     Weingarten and Challifour \cite{WeCh} formulated QED$_2$ on a finite 
lattice and showed that a Salam--Matthews formula for Schwinger functions
analogous to (\ref{MS}) holds, where the reference measure is that of
the lattice gauge theory discussed in Section \ref{YM} with gauge group
$G =$ U(1) (which gives a lattice approximation to the action of free 
Euclidean electromagnetic fields expressed in terms of the 
electromagnetic potentials), the product of scalar fields is replaced by a 
polynomial of electromagnetic potentials and the renormalized determinant is 
replaced by a similar expression. They showed \cite{WeCh,We} that in the limit 
as the lattice spacing goes to 0 and the volume cutoff is removed, their 
expressions for the Schwinger functions involving arbitrary products of the 
gauge potentials and products of pairs of fermionic and antifermonic fields
have well defined limits. However, none of the Osterwalder--Schrader axioms 
were addressed. Seiler \cite{Se2} has given a sketch of a cluster expansion 
whose convergence would verify all the Osterwalder--Schrader axioms for the 
Schwinger functions involving gauge invariant local fields (such as the 
electromagnetic field and the fermion current) plus a mass gap for the 
fermion (electron) for sufficiently small ratio of the electric charge to the 
fermion mass, but the details have not appeared in print.

\subsubsection{Yang--Mills$_d$ models} \label{YM}

     There are many more or less equivalent formulations of the classical 
Yang--Mills theory. The most concise, which requires familiarity
with some basic concepts of differential geometry, is the following: Let 
$G$ be a compact group, $A$ be a connection on a $G$-bundle over $\RR^d$ 
($A$ is a 1-form taking values in the Lie algebra $\mathfrak{G}$ of $G$),
and $F = dA + A \wedge A$ be the corresponding curvature. The (classical) 
action for the pure Yang--Mills$_d$ model is
$$\frac{1}{4g^2} \int_{\RR^d} \, \textnormal{Tr} F \wedge *F \, ,$$
where $\textnormal{Tr}$ denotes an invariant quadratic form on $\mathfrak{G}$
and $g$ plays the role of a coupling constant. Introducing coordinates on 
$\RR^d$ and a basis in $\mathfrak{G}$, one has
$$F_{\mu\nu,a} = \partial_\mu A_{\nu,a} - \partial_\nu A_{\mu,a} - 
c^{bc}_a A_{\mu,b}A_{\nu,c} \, ,$$
where $c^{bc}_a$ are the structure constants of $\mathfrak{G}$. In these terms
the action is
$$ -\frac{1}{4g^2} \int_{\RR^d} \, \sum_a \, F_{\mu\nu,a}(x) F^{\mu\nu}_a(x) \; dx^d 
\, .$$
If $G = U(1)$, then the structure constants are 0 and one recovers
the action of pure electromagnetism. In Yang--Mills theories $G$ is understood
to be nonabelian. For example, in QCD one takes $G = SU(3)$ and in the 
electroweak theory of Glashow, Salam and Weinberg one takes 
$G = SU(2) \times U(1)$ (both of these are components of the SM). In QCD and 
the SM there are, in addition, matter fields which are coupled to the gauge 
potentials $A$ and provide a further contribution to the action. These are not 
discussed here, with one exception, since the rigorous results in these cases 
are minimal at this point in time. However, there are results of interest
concerning pure Yang--Mills models, and we turn to those next. 

     For spacetime dimension $ d > 2$ one is faced again with ultraviolet
problems that go beyond Wick ordering. And since almost all means of 
regularization (\ie suppressing high enery--momentum values) destroy gauge 
invariance, most rigorous studies of (Euclidean) Yang--Mills in higher 
dimensions commence with the theory on a lattice. There are various versions 
of this, but we will consider only one, due essentially to Wilson \cite{Wil}. 

     Let $\Lambda$ be a hypercubical lattice in $\RR^d$ with lattice spacing
$a$. Ordered pairs $xy$ of nearest-neighbor lattice points $x,y \in \Lambda$
are called bonds, closed loops $uvxy$ consisting of the obvious four bonds 
are called plaquettes. Let $G$ be a compact Lie group, let $\chi$ be a 
character on $G$, and let $g_\cdot$ be a map from the bonds in $\Lambda$ into
$G$ such that $g_{xy} = g_{yx}^{-1}$. The collection of such maps is called the
field configuration space. To each plaquette $P = uvxy$ in $\Lambda$
corresponds the conjugacy class $g_P$ of the element $g_{uv}g_{vx}g_{xy}g_{yu}$
and the quantity $A_P = \frac{1}{2}(\chi(g_P) + \overline{\chi(g_P)})$.
The Wilson action\footnote{Another, less commonly used action for lattice 
gauge theories is the Villain action, which is an approximation to the Wilson 
action. Its description is lengthier and is not presented here.} on the 
lattice $\Lambda$ is defined in terms of these quantities:
$$A^W_\Lambda \equiv \frac{1}{g^2} \sum_\Lambda A_P \, ,$$
where the sum runs over all plaquettes $P$ in $\Lambda$. A probability measure
on the field configuration space is given by
$$\mu_\Lambda \equiv 
\frac{e^{-A^{W}_\Lambda} \prod dg_{xy}}{\int e^{-A^{W}_\Lambda} \prod dg_{xy}} \, ,$$
where the product is taken over all bonds $xy$ in the lattice and
$dg_{xy}$ is the Haar measure on $G$ for each such bond. Gauge transformations
are given by $g_{xy} \mapsto \gamma_x g_{xy} \gamma_y^{-1}$, where 
$\gamma : \Lambda \rightarrow G$. Both the action and the measure are
gauge invariant. As shown by Osterwalder and Seiler \cite{OsSei}, the 
expectations with respect to $\mu_\Lambda$ of gauge invariant functions on the 
field configuration space satisfy a natural analogue of the reflection 
positivity condition in the Osterwalder--Schrader axioms, so that one
can construct a corresponding Hilbert space and a positive self-adjoint 
transfer matrix (also bosons and fermions can be naturally incorporated
into this setting \cite{OsSei}). Much rigorous work has treated the 
infinite volume limit of such theories, but for the purposes of quantum
field theory one also must control the limit as $a \rightarrow 0$. 
For this, there are significantly fewer results.

     Gross \cite{Gr} showed that as the lattice spacing converges to 0 
the U(1)$_3$ lattice gauge model with the Villain action converges to the free 
Euclidean electromagnetic field in the sense of convergence of the 
characteristic functions of the field variables $F_{\mu\nu}$. He also showed
that if the Wilson action is adopted, then the characteristic function
of the ``lattice current'' converges to the characteristic function of the
current of the free Euclidean electromagnetic field as the lattice spacing
goes to 0. Driver \cite{Dri} proved that similar results hold for the
U(1)$_4$ lattice gauge model.

     In two spacetime dimensions, when the complete axial gauge is 
taken the Euclidean Yang--Mills action simplifies sufficiently to define
a Gaussian measure for the gauge field. This greatly reduces the difficulties
in the task of construction and affords the 
possibility of avoiding lattice approximations (though at the cost of
giving up explicit gauge invariance), and a number of different approaches 
have been developed to that end. Only a few highlights can be mentioned here. 
Gross, King and Sengupta \cite{GrKiSe} constructed the Euclidean U($N$) pure 
Yang--Mills model in the axial gauge in two spacetime dimensions using 
stochastic differential equations (\cf Section \ref{stoch}). They show that 
the differential equation controlling parallel transport along a smooth curve 
in $\RR^2$ with respect to a typical gauge potential can be interpreted as a 
stochastic differential equation. Without cutoffs they construct the Schwinger 
functions for the (nonoverlapping) Wilson loops, provide closed expressions 
for these Schwinger functions and verify their Euclidean invariance. Driver 
\cite{Dr} extended their results to include expectations of more general 
functions of parallel transport along a finite ``admissible collection'' of 
curves and further proved that the continuum model is the limit of the (gauge 
fixed) lattice approximations (for both the Villain and the Wilson actions) as 
the lattice spacing goes to 0. Anshelevich and Sengupta \cite{AnSe} have
shown that the limit $N \rightarrow \infty$ (with $N/g^2$ held fixed) of
these Euclidean U(N) pure YM$_2$ models exists and is described by a 
concrete free stochastic process.

     These results have been extended in many respects by Ashtekar and 
co-authors \cite{ALMMT}. Studying Euclidean pure YM$_2$ models with $G$ equal 
to SU($N$) or U(1), they take the reference measure $\mu_0$ on a 
suitable closure of $\mathscr{A}/\mathscr{G}$ discussed above, perturb it by 
$\exp\{ -A^W_\Lambda \}$ (see below), normalize to obtain a probability measure, 
and explicitly compute the expectations of products of Wilson loops. They then 
show that the resulting expressions have a well defined limit as the volume 
cutoff is removed and the lattice spacing goes to 0.\footnote{Fleischhack 
\cite{Fl} has repaired some technical problems in their work, recovering their 
results.} They therefore have rigorously established a closed expression for 
the expectations of products of generic Wilson loops for YM$_2$. In addition, 
they explicitly construct the corresponding real time model and establish the 
equivalence between the Euclidean and real time formulations. Klimek and 
Kondracki \cite{KlKo} have constructed a measure without cutoffs for Euclidean 
YM$_2$ with $G =$ SU(2) coupled to a massive fermion, but they did not verify 
any of the usual axioms. 

     There has been progress in the rigorous analysis of YM$_4$, primarily
for $G =$ SU(N), though in the few places where the details of the group make
a difference in the proof of the estimates $G=$ SU(2) is usually taken. In a 
highly technical argument incorporating renormalization group transformations 
and stretching over several lengthy papers, Balaban (see \cite{Bal2} for 
references) succeeded in proving the ultraviolet stability of the model on a 
lattice of arbitrary bond length. This should be the core of 
a proof that the ultraviolet limit (here in the guise of the lattice spacing 
going to 0) of the expectations of products of Wilson loops exists. However, 
the series of papers came to an end before exact Schwinger functions were 
constructed and their properties were verified. Unfortunately, the same is 
true of a series of papers by Federbush on the same model (see \cite{Fe2}
for references). 

     The problem of constructing a pure Yang--Mills model in four spacetime 
dimensions remains open. In fact, the Clay Mathematics Institute has offered 
one million dollars to anyone who succeeds in showing that for any compact 
simple group $G$ the corresponding pure Yang--Mills model in four spacetime 
dimensions exists, satisfies conditions at least as strong as the Wightman or 
Osterwalder--Schrader axioms, is nontrivial and manifests a mass gap.

\section{Functional Integral Constructions --- Minkowski}  \label{path2}
\setcounter{equation}{0}

     Real time functional integrals are technically more difficult to
work with than the corresponding Euclidean integrals, since they are 
oscillatory in nature and do not benefit from exponential suppression of
large ranges of field ``values'', as do the Euclidean integrals. Although 
much progress has been made in the control of the functional integrals arising 
in nonrelativistic quantum mechanics, the situation for the real time 
functional integrals that are relevant to QFT is much less developed.
The reader is referred to the monograph of Klauder \cite{Kl1} for an
accounting of both. There is not yet sufficient progress to treat interacting 
relativistic quantum fields rigorously in this framework.

\section{Constructions Using Stochastic Differential Equations} \label{stoch}
\setcounter{equation}{0}

     The strong links between Euclidean QFT and probability theory have been 
suggested in Section \ref{path1}, but we discuss another such link in this 
section. If $B_t$ is a cylindric version of $\Ss'(\RR^{d-1})$-valued Brownian
motion, so that for each $f \in \Ss(\RR^{d-1})$ $b_t^f \equiv B_t(f)$ is
a version of one dimensional Brownian motion, then the linear Ito
stochastic differential equation (SDE)
\begin{equation}  \label{ito}
d\xi_t^0 = \sqrt{-\Delta_{d-1} + 1} \, \xi_t^0 \, dt + dB_t
\end{equation}
has a stationary Gaussian solution $\xi_t^0$ with mean 0 and expectation which
coincides with that of the free Euclidean scalar Bose field with mass 1:
$$ E \xi_0^0({\bf x}) \xi_t^0({\bf y}) = 
\int_{\Ss'(\RR^d)} \, \phi(0,{\bf x}) \phi(t,{\bf y}) \; d\mu_{C_1} \, .$$
These stochastic processes can therefore be identified in the indicated sense.
Another such relation of note is provided by considering a Gaussian white
noise $\eta$ on $\Ss'(\RR^d)$, \ie $\eta$ is a $\Ss'(\RR^d)$-valued
random variable distributed according to a Gaussian measure $\nu$
with generating function
$$\int_{\Ss'(\RR^d)} \; e^{i \eta(f)} \, d\nu(\eta) = 
\exp(-\frac{1}{2} \Vert f \Vert_2^2) \, .$$
For $\lambda \in (0,\frac{1}{2}]$, the solution of the stochastic partial
(pseudo-)differential equation (SPDE) 
$$(-\Delta + 1)^\lambda \phi_\lambda = \eta $$
is given by the stochastic convolution integral
$$\phi_\lambda = (-\Delta + 1)^{-\lambda} * \eta \, .$$
$\phi_\lambda$ is a Gaussian stochastic process of mean 0 and covariance
$$E \phi_\lambda(x) \phi_\lambda(y) = (-\Delta + 1)^{-2\lambda}(x-y) \, ,$$
so that $\phi_{1/2}$ can be identified with the free Euclidean scalar Bose 
field with mass 1. 

     To get beyond Gaussian processes (and thus free quantum
fields), essentially two approaches have been introduced. One is to perturb the 
linear drift term in (\ref{ito}) with a nonlinear term. In this case the typical
sample paths are distributions instead of functions. The second is to change
the reference Gaussian process (Brownian motion or white noise) into a
suitable non-Gaussian process, for example to perturb the white noise
with Poisson noise. In both one typically finds a stochastic process which
is a weak solution of the SDE one has set up; then the equilibrium measure 
corresponding to this process is the measure sought in the Euclidean QFT
program.

     These methods or their variants have been employed by many authors for 
the purpose of constructing quantum field models. As an example, we summarize
work of Jona-Lasinio and Mitter \cite{JoMi}. Let $\Lambda$ be a square region 
in $\RR^2$, $\Delta_\Lambda$ be the Laplace operator with Dirichlet boundary 
conditions on $\Lambda$ and $C = (-\Delta_\Lambda + 1)^{-1}$. For 
$\varepsilon \in (0,\frac{1}{10})$ consider the nonlinear SDE
\begin{equation}  \label{ppp}
d\hat{\phi}_t = -\frac{1}{2}(C^{-\varepsilon} \hat{\phi}_t + 
\lambda C^{1-\varepsilon}\, : \hat{\phi}_t^3 :)\, dt + dW_t 
\end{equation}
with some specified initial condition $\iota$,\footnote{The solution and its 
laws will therefore depend on the choice of $\iota$.} where $W_t$ is a Brownian 
motion with mean 0 and covariance
$$E \, W_t(f)W_s(g) = \langle f, C^{1-\varepsilon}g \rangle \, \min\{ t,s \} \, .$$
Jona-Lasinio and Mitter construct a Markov process $\hat{\phi}_t$ which is
a weak solution of (\ref{ppp}), is ergodic and mixing, and satisfies
$$\lim_{t \rightarrow \infty} E_{\hat{\phi}_0 = \iota} \, 
\hat{\phi}_t(f_1) \ldots \hat{\phi}_t(f_n) = 
\int \; \phi(f_1) \ldots \phi(f_n) \; d\mu_\Lambda(\phi) \, ,$$
where $\mu_\Lambda$ is the measure given in (\ref{pp}) with $P(\phi) = \phi^4$
and $C$ the covariance mentioned directly above. Hence, they have constructed, 
in this sense, the $\phi^4_2$ model in a finite volume. Similar constructions 
have been made of the $P(\phi)_2$ models, the sine--Gordon model (in the range 
$\alpha^2 < 4\pi$) and the exponential interaction model in two spacetime 
dimensions, all with a volume cutoff (see \eg \cite{Wu}). 

     The process $\hat{\phi}_t$ constructed in this manner is 
$\Lambda$-dependent. Of particular interest is the fact that Borkar, Chari and 
Mitter \cite{BoChMi} showed that one can take the infinite volume limit also 
on the left hand side of the above equality (the processes indexed by 
$\Lambda$ converge in a suitable sense as $\Lambda \nearrow \RR^2$). Since it 
is already known that the limit of the right hand side as 
$\Lambda \nearrow \RR^2$) yields the Schwinger functions of the fully 
interacting $\phi^4_2$ model, one sees that this approach also succeeds in 
constructing the exact $\phi^4_2$ model. The methods of Borkar, Chari and Mitter
should also be applicable to general $P(\phi)_2$ models.

     Some effort has been expended to understand renormalization from the
point of view of stochastic quantization. However, this has not yet resulted 
in a construction of a quantum field model requiring an ultraviolet 
renormalization going beyond Wick ordering. 

     Albeverio and co-workers have constructed a class of Euclidean random
fields via convolution from generalized white noise and have shown that
the Schwinger functions of these models can be analytically continued
to real time ``Wightman functions'' (\cf \cite{AlGoWu} for details and
references). All of the Wightman axioms hold for these models except possibly
the positivity condition\footnote{In some instances it has been proven that the
positivity condition is, in fact, violated.} on the family of Wightman 
functions that allows one to construct a Hilbert space in which the field 
operators act. Albeverio, Gottschalk and Wu \cite{AlGoWu} have shown that 
their ``Wightman functions'' satisfy a weakened condition due to Morchio and 
Strocchi (\cf \cite{Str}) motivated by the study of local gauge theories 
(see Section \ref{gauge}). This condition assures that one can construct a 
Krein space, instead of a Hilbert space, in which the special subset $\Ds$ 
mentioned in Section 1 is dense. In these models one may choose $d$ as one 
likes, so this approach results in QFT models in four spacetime dimensions 
represented in Krein spaces, instead of Hilbert spaces. The scattering behavior 
in such models has also been studied by Albeverio and Gottschalk \cite{AlGo}, 
and some of the models manifest nontrivial scattering. However, it is not clear 
at this point whether these models include any of the models of interest in 
elementary particle physics.

\section{\hspace*{-5mm} Algebraic Constructions II}  \label{algebra2}     
\setcounter{equation}{0}

     Relatively recently, new insights and tools attained in AQFT have
led to an important renaissance of algebraic real time constructions of
quantum field models, resulting in new techniques and the construction of
models which either cannot be constructed by other known techniques
or can only be so constructed with a prohibitive amount of effort. In
these constructions one is not guided by Lagrangian QFT but rather
uses other input to arrive at the models.

     For many of these constructions a particular collection of
spacetime regions is distinguished --- the so-called wedges.
After choosing an inertial frame of reference in Minkowski space, one defines 
the right wedge to be the set \newline
$\Ws_R = \{ x = (x_0,x_1,x_2,x_3) \in \RR^4 \mid x_1 > \vert x_0 \vert \}$
and the set of wedges to be
$\bcW = \{ \lambda \Ws_R \mid \lambda \in \Pid \}$.
The set of wedges is independent of the choice of reference frame;
only which wedge is designated the right wedge is frame-dependent.
Let $\theta_{\Ws}$ denote the reflection on Minkowski space about the
``edge'' of the wedge $\Ws$. The set $\{ \theta_{\Ws} \mid \Ws \in \bcW \}$
generates the proper Poincar\'e group $\Ps_+$. 
These regions have been of interest since an important insight won by
Bisognano and Wichmann \cite{BiWi}, namely that for any net $\net$
of von Neumann algebras locally associated to a Wightman theory 
$(\phi, \Hs, U, \Omega)$ in a natural manner\footnote{Roughly speaking,
the bounded functions of the field operators $\phi(f)$ smeared with
test functions $f$ having support contained in the spacetime region $\Os$
are contained in the algebra $\As(\Os)$.} the abstract modular objects 
$\Delta_\Ws$, $J_\Ws$ associated with the pair $(\As(\Ws),\Omega)$ by 
Tomita--Takesaki theory \cite{Ta2,BrRo,Su3} have, in fact, physical meaning. 
They showed that one has
\begin{equation} \label{modular1}
J_{\Ws_R} = \Theta \, U(R_1(\pi)) \, , \, 
\Delta_{\Ws_R}^{it} = U(v_R(2\pi t)) \, , 
\end{equation}
where $\Theta$ is the PCT-operator associated to the Wightman field,
$v_R(t)$, $t \in \RR$, is the one-parameter subgroup of boosts leaving the
wedge $\Ws_R$ invariant, and $R_1(\pi)$ is the rotation through the angle 
$\pi$ about the $1$-axis, which is perpendicular to the edge of $\Ws_R$
(similar relations are valid for the modular objects of any wedge). Hence, 
$$J_{\Ws_R} \As(\Os) J_{\Ws_R} = \As(\theta_{\Ws_R} \Os) \, , \, 
\Delta_{\Ws_R}^{it} \As(\Os) \Delta_{\Ws_R}^{-it} = 
\As(v_R(2\pi t) \Os) \, ,$$
for all $\Os$. Note for later use that 
$\Delta_{\Ws_R}^{it} = U(v_R(2\pi t)) = e^{i2\pi tK_1}$ entails 
\begin{equation} \label{modular2}
\Delta_{\Ws_R}^{1/2} = e^{\pi K_1} \, . 
\end{equation}
Though some of the significance of their insight will become clear below, we 
refer the reader to \cite{Su4} for further development and references.

     A fairly general strategy \cite{BaWo,BuSu2} to construct a quantum field 
theory algebraically is to construct a unitary representation $U(\Pid)$ (the 
representation would be fixed by an analysis of particle masses, types and 
multiplicities in scattering experiments) satisfying the spectrum 
condition\footnote{Indeed, due to the fact that these wedge algebras
are generally type III factors and therefore unitarily equivalent, it would 
suffice to do so on Fock space.} and for a fixed wedge $\Ws_0$ exhibit an 
algebra ${\mathfrak G}$ which satisfies the {\it consistency conditions}:

   (a) $U(\lambda) \, {\mathfrak G} \, U(\lambda)^{-1} \subset {\mathfrak G}$,
whenever $\lambda \Ws_0 \subset \Ws_0$ for $\lambda \in \Pid$.

   (b)  $U(\lambda^\prime) \, {\mathfrak G} \, U(\lambda^\prime)^{-1} \subset 
{\mathfrak G}^\prime$, whenever $\lambda^\prime \, \Ws_0 \subset \Ws_0^\prime$ 
for $\lambda^\prime \in \Pid$, where a primed algebra denotes the commutant
of the algebra and a primed wedge denotes the complementary wedge, \ie
the maximal wedge which is spacelike separated from the given wedge. 

     Then setting 
$\As(\Ws) \equiv U(\lambda) \, {\mathfrak G} \, U(\lambda)^{-1}$,  
where $\lambda \in \Pid$ is such that $\Ws = \lambda \Ws_0$ for any 
$\Ws \in \bcW$, one obtains an ``algebra of observables'' for each wedge
region. For any other (convex) causally complete spacetime region $\Os$ one 
can then define $\As(\Os)$ as the intersection of all wedge algebras $\As(\Ws)$
such that $\Os \subset \Ws$. The resulting net $\net$ satisfies the
conditions of isotony, relativistic covariance and Einstein causality. 
Conversely, any asymptotically complete quantum field theory fixes an algebra 
${\mathfrak G}$ satisfying the consistency conditions. One may therefore
view such pairs $(U,{\mathfrak G})$ as germs of quantum field models.
However, at present one does not have a general dynamical principle by which 
the algebras ${\mathfrak G}$ may be selected, given the representation $U$. 
In the following, we briefly outline what may be seen as various ways of 
arriving at such pairs.

\subsection{Modular localization}

     Though there have been many constructions of free quantum fields
using various techniques, Bisognano and Wichmann's breakthrough motivated
Brunetti, Guido and Longo \cite{BrGuLo} to find what may be viewed as an 
intrinsic construction which we discuss next. We restrict our discussion to
$d = 4$; the case $d = 3$ is more complex, as anyons are admitted 
\cite{MuSchYn}. One of the manifold aspects of 
the subtle notion of ``particle'' in QFT arises from Wigner's famous 
classification of the irreducible representations of the (covering group of 
the) Poincar\'e group \cite{Wi}, in which each such representation is uniquely 
determined (up to a natural equivalence) by two numbers that can be 
interpreted as the mass and the spin of the particle (if the mass is strictly 
positive; when the mass is zero, there is the additional case of ``continuous
spin''). So presented with the task to construct a
quantum field describing free particles of given mass and spin, one
begins with the corresponding irreducible representation
$U_1(\Ps_+)$ of the proper Poincar\'e group on a Hilbert space $\Hs_1$.
From this data one constructs a representation $U(\Ps_+)$ on $\Hs$, 
the bosonic Fock space with one--particle space $\Hs_1$.
Writing the representation of the boost subgroup leaving $\Ws_R$ invariant 
in terms of its self-adjoint generator, $U_1(v_R(t)) = e^{itK_1}$, one 
{\it defines}, in light of (\ref{modular1}) and (\ref{modular2}) above,
$$\Delta_{\Ws_R}^{1/2} \equiv e^{\pi K_1} \, , \, 
J_{\Ws_R} \equiv U_1(\theta_{\Ws_R}) $$
Motivated by Tomita--Takesaki theory one then defines
$$S_{\Ws_R} \equiv U_1(\theta_{\Ws_R}) e^{\pi K_1} = J_{\Ws_R} \Delta_{\Ws_R}^{1/2}$$
and identifies the corresponding real subspace of invariant vectors in
$\Hs_1$
$$\Ks_{\Ws_R} \equiv \{ f \in D(S_{\Ws_R}) \mid S_{\Ws_R} f = f \}$$
(similarly for all $\Ws \in \bcW$). On the exponential (or {\it coherent}) 
vectors 
$$e^h \doteq \oplus_{n = 0}^{\infty} \frac{1}{\sqrt{n!}} \, h^{\otimes n } \, ,
\, h \in \Hs_1 \, , $$
which are total in $\Hs$, one defines unitary operators $V(f)$, $f \in \Hs_1$, 
by
$$V(f) e^0 \doteq e^{- \frac{1}{4} \Vert f \Vert^2} e^{\frac{i}{\sqrt{2}} f} 
\, , \, f \in \Hs_1 \, , $$
$$V(f) V(g) \doteq e^{- \frac{i}{2} \textnormal{Im} \langle f , g \rangle} V(f+g) \, ,
\, f,g \in \Hs_1 \, . $$
Now let $\As(\Ws_R)$ denote the von Neumann algebra generated by
$\{ V(f) \mid f \in \Ks_{\Ws_R} \}$. Brunetti, Guido and Longo showed that
if $U_1(\Ps_+)$ satisfies the spectrum condition, then $(U(\Pid),\As(\Ws_R))$ 
satisfies the consistency conditions and therefore determines a local, 
Poincar\'e covariant net $\{ \As(\Os) \}$. In the case of finite spin,
the resulting net coincides with the net of local algebras associated
with the corresponding free field; in the special case of massless,
``infinite (or continuous) spin'' representations $U_1$, this provides the 
first construction of a quantum field model covariant under an infinite spin 
representation. Indeed, there is no quantum field of the standard type
associated with the net in this case, since the Fock vacuum vector 
$\Omega = e^0$ is not cyclic for algebras $\As(\Os)$ when $\Os$ is bounded;
instead, these nets are generated by string-localized fields \cite{MuSchYn}.
However, for any choice of positive energy representation $U_1$ the vector 
$\Omega$ is cyclic for $\As(\Ws)$, for any wedge $\Ws$, and for 
$\As(\Cs)$, for any spacelike cone $\Cs$. In the case that $U_1(\Ps_+)$ is 
irreducible with finite spin, $\Omega$ is also cyclic for $\As(\Os)$, 
for any double cone $\Os$.

\subsection{Models with nontrivial factorizing S--matrices} \label{factor}

     A classic problem of obvious physical importance is the so-called
{\it inverse scattering problem}, which in the context of QFT is:
given the scattering matrix $S$ (determined, in principle, by the measured 
scattering data in an experiment), does there exist a quantum field model 
whose associated S--matrix is the specified operator $S$? As this is an enormously
difficult problem, in order to make the problem more manageable workers in 
the field have focused their attention on the special case of factorizing 
S--matrices in two spacetime dimensions; this is the relatively simple 
situation where all scattering processes reduce to (suitable combinations of) 
two-body scattering, so that specification of the two-body scattering 
amplitude completely determines the S--matrix. There is, therefore, no
particle production in such models. The primary effort in this 
direction has been made in the context of the form factor program 
\cite{KaWe,ZaZa,Sm}, in which local quantum field operators associated with 
the quantum field purportedly having the prescribed scattering behavior are 
expressed in terms of a certain algebra, the Zamalodchikov algebra. Rigorous 
formulas for matrix elements of local operators between scattering states have 
been obtained, but the computation of products of local operators at different 
spacetime points is not under mathematical control, because infinite sums over 
intermediate states are involved. Hence, the Wightman axioms have not been 
verified in such models, apart from simple cases.   

     However, Schroer \cite{Schr1,Schr2} realized that certain field operators 
in the Zamalodchikov algebra can be interpreted as being localized in wedges. 
These wedge-localized but nonlocal quantum field operators that create 
covariant one-particle states out of the vacuum are now called
{\it polarization-free generators}. It was subsequently shown by Borchers,
Buchholz and Schroer \cite{BoBuSch} that in more than two spacetime
dimensions the existence of (tempered) polarization-free generators defined 
on a translation-invariant, common dense domain (tacitly assumed in 
\cite{Schr1,Schr2}) entails the triviality of the associated S--matrix. 
However, for constructing quantum field models with nontrivial scattering in 
two dimensional Minkowski space the idea is quite fruitful.

     In the special case of these factorizing S--matrix models the 
S--matrix is determined by a single function $S_2$ through the relation 
$$(S \Psi)_n(\theta_1,\ldots,\theta_n) = 
\big[ \prod_{1 \leq l < k \leq n} S_2(\vert \theta_l - \theta_k \vert) \big]
\Psi_n(\theta_1,\ldots,\theta_n) \, , $$
which defines its action on a general $n$-particle wave function in the Hilbert
space described below, where $\theta_i$, $i = 1,\ldots,n$, are the rapidities 
of the scattered particles. Lechner \cite{Le1,Le2,Le3} considers a large class 
of two body scattering functions $S_2(\theta)$ satisfying conditions arising 
from the requirements of unitarity, crossing symmetry and hermitian analyticity 
of the corresponding S--matrix and uses them to define a concrete 
representation of the Zamalodchikov algebra. With $\Hs_1 = L^2(\RR)$, let 
$\Hs$ be the $S_2$--symmetrized Fock space, where the $n$-particle wave 
functions satisfy
$$\Psi_n(\theta_1,\ldots,\theta_{i+1},\theta_i,\ldots,\theta_n) =
S_2(\theta_i - \theta_{i+1})
\Psi_n(\theta_1,\ldots,\theta_{i},\theta_{i+1},\ldots,\theta_n) \, .$$
For the special, and admissible, scattering functions $S_2 = 1, -1$, $\Hs$
is the standard boson, respectively fermion, Fock space. $\Hs$ admits a
canonical, strongly continuous unitary representation $U(\Pid)$ of the 
Poincar\'e group satisfying the spectrum condition. On $\Hs$ also act 
creation, resp. annihilation, operators $Z^\dagger$, $Z$, satisfying the 
Fadeev--Zamolodchikov relations:
$$Z^\dagger(\theta)Z^\dagger(\theta') = S_2(\theta - \theta')
Z^\dagger(\theta')Z^\dagger(\theta) $$
(similarly for $Z$) and
$$Z(\theta)Z^\dagger(\theta') = S_2(\theta' - \theta) 
Z^\dagger(\theta')Z(\theta) + \delta(\theta - \theta') \cdot \idty$$
In terms of these one defines a quantum field operator
$$\phi(f) \doteq Z^\dagger(f_+) + Z(f_-) \, , \, f \in \Ss(\RR^2) \, ,$$
where
$$f_\pm(\theta) = \frac{1}{2\pi} \int \, f(x) e^{\pm i p(\theta)x} \, dx $$
and $p(\theta) = m(\cosh \theta, \sinh \theta)$, where $m > 0$ is the mass.

     Though the field $\phi(x)$ is covariant under the action of $U(\Pid)$,
is densely defined, and is a distributional solution of the Klein-Gordon
equation of mass $m$, it does not satisfy Einstein causality unless
$S_2 = 1$, in which case $\phi(x)$ is the usual free scalar Bose field.
However, as observed by Schroer, when smeared with test functions having
support in a wedge, they are polarization-free generators.   

    In two dimensional Minkowski space the set of wedges decomposes into
two components, one is the set of all translates of $\Ws_R$ and the other
is the set of all translates of $\Ws_L \equiv \Ws_R^\prime$. Defining 
$\As(\Ws_R)$ to be the von Neumann algebra generated by 
$\{ e^{i\phi(f)} \mid \supp(f) \subset \Ws_R \}$, 
$\As(\Ws_L) \equiv \As(\Ws_R)^\prime$, and 
$\As(\Ws) \equiv U(x) \As(\Ws_R) U(x)^{-1}$ if $\Ws = \Ws_R + x$ (similarly
for $\Ws = \Ws_L + x$), then $\wnet$ satisfies the HAK axioms for the
restricted collection $\Rs = \bcW$. The Fock vacuum vector $\Omega$ is
cyclic for the wedge algebras and the corresponding modular objects
coincide with those found in the Bisognano--Wichmann setting. 

     In two spacetime dimensions double cones can be defined simply as 
the intersection of two suitable wedges: $\Os = \Ws_1 \cap \Ws_2^\prime$
with $\Ws_2 \subset \Ws_1$. Lechner showed that if one defines the
corresponding observable algebra to be
$$\As(\Os) \equiv \As(\Ws_1) \cap \As(\Ws_2)^\prime \;
( = \As(\Ws_1) \cap \As(\Ws_2^\prime)) \, ,$$
one finds that for a large class of $S_2$ (\ie for a large class of models of
the type we are discussing), $\Omega$ is cyclic and separating 
for all double cone algebras and the double cone algebras satisfy Einstein
causality. Moreover, the Haag--Ruelle scattering theory can then be applied 
to yield a scattering theory which is asymptotically complete and
whose S--matrix coincides with the initially prescribed S--matrix \cite{Le3}.
These models therefore constitute a complete and satisfying solution to the
inverse scattering problem for the stated class of S--matrices.  

     Note that the construction is implemented using easily constructed,
but nonlocal fields to obtain local wedge algebras, of which suitable relative
commutants provide algebras of observables localized in bounded
spacetime regions. This sidesteps the usual process
of first constructing local fields and then using them to obtain the local
observable algebras. This is a significant simplification for the models
in question, since arguments by Smirnov and Schroer \cite{Sm,Schr2} and 
examples laboriously computed by McCoy {\it et alia} \cite{MTW} indicate 
that the {\it local} fields in these models must be {\it infinite power series} 
in the (simple) nonlocal fields. By constructing the local algebras in the 
indicated manner, one is able to avoid controlling the infinite expansion 
that yields the local fields and yet still arrive at the desired quantities of
physical relevance.

     The special case of $S_2 = -1$ was studied by Buchholz and Summers 
\cite{BuSu1} for $d \geq 2$, and it was shown that the model is maximally 
nonlocal in a certain specific quantitative sense. For $d = 2$ it was 
shown that there are two associated local nets admitting an asymptotically 
complete scattering theory, one describing a fermion with trivial scattering 
and another describing a boson with $S = (-1)^{N(N-1)/2}$, where $N$ is the 
number operator. In higher dimensions there exist string-localized, 
respectively brane-localized, operators which mutually commute at spacelike 
separation. \cite{BuSu1}

\subsection{Deformations} \label{warp}

     Deformation techniques have long been used to provide quantized versions
of classical theories (see \eg \cite{Wa}), but attempts to deform quantum
field models have appeared fairly recently. Their initial impetus was provided
by the desire to deform quantum fields on Minkowski space to arrive at
quantum fields on Moyal space, or noncommutative Minkowski space. Though most 
of that work has not been mathematically rigorous, Grosse and Lechner 
\cite{GrLe} rigorously defined a deformation of the scalar massive free Bose 
field in Wightman's setting, which could then be interpreted either as a model 
on noncommutative Minkowski space or as a model on Minkowski space. Of 
particular interest is that the deformed field manifests nontrivial scattering 
(see below). Buchholz and Summers \cite{BuSu2} then found a deformation of 
essentially any algebraic quantum field model which coincides with the 
deformation of Grosse and Lechner when restricted to the net associated with 
the mentioned free field. We discuss this deformation and the resulting models 
here. Though this procedure may be carried out for $d \geq 2$, we restrict our 
discussion to $d = 4$. 

     Let $(\net, \Hs, U, \Omega)$ satisfy the conditions stated in Section
\ref{intro}; let $U(x) = e^{iPx}$, $x \in \RR^4$, and $P = \int p \; dE(p)$
be the joint spectral decomposition of the generators of $U(\RR^4)$.
The support of the projection-valued measure $E$ is $\overline{V_+}$,
by the spectrum condition. Let $M$ be a bounded operator and, for convenience,
let
$$ \alpha_{p}(M) \equiv e^{iPp} M e^{-iPp} \in \Bs(\Hs) \, , p \in \RR^4 \, .$$
The matrices
\begin{equation*}  \label{q}
Q \equiv 
\left( \begin{array}{cccc}  0 & \zeta & 0 & 0 \\ 
                          \zeta   &  0 & 0 & 0 \\
                            0 &    0   & 0 & \eta \\
                            0 &    0   & -\eta & 0 \end{array} \right)
\end{equation*}
for fixed $\zeta \geq 0$, $\eta \in \RR$, are uniquely distinguished by the 
following properties \cite{GrLe}:
\begin{enumerate}
\item[(i)] \ $Q \, \overline{V_+} \subset \Ws_R $. 
\item[(ii)] 
\ If $\lambda = (\Lambda,x) \in \Pid$ satisfies $\lambda \Ws_R \subset
\Ws_R $, then $\Lambda Q \Lambda^T = Q$. 
\item[(iii)] 
\ If $\lambda = (\Lambda,x) \! \in \! \Pid$ satisfies 
$\lambda \, \Ws_R \! \subset \! \Ws_R^\prime $, then 
$\Lambda Q \Lambda^{T} \! = \! - Q$.
\end{enumerate}
The {\it warped convolution} of $M$ is given by the formal 
expression\footnote{Strictly speaking, this is well defined for those bounded
$M$ which are smooth with respect to the action $\alpha_x$, $x \in \RR^4$
\cite{BuLeSu}.}
\begin{equation} \label{deform}
M_Q = \int dE(p) \, \alpha_{Qp}(M) \, , \, M \in \Bs(\Hs) \, .
\end{equation}
A detailed explication of this unusual operator valued integral is given
in \cite{BuLeSu}. One finds that $M_Q$ is also a bounded operator on $\Hs$,
so one may further define the deformed algebra $\As_Q(\Ws_R)$ corresponding to 
the algebra $\As(\Ws_R)$ to be the von Neumann algebra generated by
$\{ A_Q \mid A \in \As(\Ws_R) \}$. Then $(U,\As_Q(\Ws_R))$ satisfies the
consistency conditions \cite{BuSu2,BuLeSu} and therefore generates, as above, 
a net satisfying the Haag-Araki-Kastler axioms. Although the original net
and the deformed net are not isomorphic, the modular objects
corresponding to the pair $(\As(\Ws_R),\Omega)$ coincide with those
corresponding to $(\As_Q(\Ws_R),\Omega)$ \cite{BuLeSu}. 

     There are indications that the deformed algebras corresponding to
bounded spacetime regions $\Os$ may be trivial, \ie are multiples of
the identity operator, so one may not be able to formulate a full scattering 
theory for the deformed net. However, two-body scattering for the
deformed model is well defined \cite{BuSu2}. The relations between the 
two-body scattering states in the original and in the deformed theory are most 
transparent if one uses improper single particle states of sharp momentum 
$p = (\sqrt{\bcp^2 + m^2}, \bcp)$, $q = (\sqrt{\bcq^2 + m^2}, \bcq)$. 
There one has \cite{BuSu2}
\begin{equation*}
\begin{split}
|p \otimes_Q q \rangle^{\mbox{\scriptsize in}}
& = e^{i |p Q q |} \, |p \otimes q \rangle^{\mbox{\scriptsize in}} \\
|p \otimes_Q q \rangle^{\mbox{\scriptsize out}}
& = e^{- i |p Q q |} \, |p \otimes q \rangle^{\mbox{\scriptsize out}} \, .
\end{split}
\end{equation*}
The scattering states in the deformed theory depend on the matrix
$Q$ through the choice of the wedge $\Ws_R$ and thus break the
Lorentz symmetry in $d > 2$ dimensions.  This can be understood if one
interprets the deformed theory as living on noncommutative Minkowski space, 
where the Lorentz symmetry is broken.

     The kernels of the elastic scattering matrices in the 
deformed and undeformed theory are related by 
$${}^{out}{\langle p \otimes_Q q |p^\prime \otimes_Q q^\prime
  \rangle}{}^{in}
= e^{i|pQ q| + i|p^\prime Q q^\prime|} \; \; 
{}^{out} \langle p \otimes q | p^\prime \otimes q^\prime \rangle{}^{in} \, .
$$
Thus they differ, and even if the initial model has trivial scattering,
the deformed theory does not. Hence, this deformation applied to the net
of the scalar massive free field results in a mathematically rigorous
quantum field model with nontrivial scattering, apparently the first such 
model in four spacetime dimensions. 

     Dybalski and Tanimoto \cite{DyTa} have shown that application of
the warped convolution to any chiral conformal QFT in two spacetime dimensions
results in a model which has Lorentz invariant nontrivial scattering and is 
asymptotically complete, the first example of a massless QFT having such 
properties.

     These deformations were subsequently reinterpreted in the Wightman
framework by Grosse and Lechner \cite{GrLe2} in terms of a deformed product 
on the Borchers algebra associated with the polynomials in the field operators 
mentioned in Section 1. And, again in that framework, Lechner 
\cite{Le4} has found a larger class of deformations which includes warped 
deformations as a special case. In four spacetime dimensions, the resulting 
deformed theories have properties similar to those of the warped models; 
however, in two dimensions he showed that the models discussed in Section 
\ref{factor} are obtained from the free field by a deformation of this new 
type. Common to all of these deformed models so far is the absence of 
particle production.

     More recently, in the context of chiral QFT in two spacetime 
dimensions whose many constraints enable the use of special techniques, Tanimoto
\cite{Tan} showed that deformations based on Longo-Witten endomorphisms are
unitarily equivalent to those of the warped convolution transform, and in 
\cite{LeSchTan} Lechner, Schlemmer and Tanimoto show that in this special 
case the infinite family of deformations presented in \cite{Le4} results in
the same equivalent results.

\section{Perturbative AQFT} \label{perturb}
\setcounter{equation}{0}

     As only briefly indicated in Section 1, the algebraic approach to QFT
may comfortably consider local algebras of observables which are not
algebras of bounded operators acting on a Hilbert space. In this section
we discuss briefly an interesting example of this which goes well beyond
the polynomial algebras of the Wightman approach. This work, carried out
primarily by Brunetti, D\"utsch and Fredenhagen, explicitly incorporates 
perturbative QFT into the framework of AQFT. Though one of the primary recent 
concerns of these authors is to formulate perturbative AQFT in a manner 
independent of any background space--time, we shall present an early version 
of their formulation, since we are concerned here with relativistic QFT on 
Minkowski space. This will simplify the discussion somewhat. For the same 
reason, we shall not try to describe the steps taken by the authors to 
incorporate more general interactions and shall, instead, sketch an early 
version of this approach.  

     As with Borchers algebras, the connection between an abstract 
(complex, unital) *-algebra $\As$ and operators acting on a Hilbert space 
$\Hs$ (where the probabilistic interpretation familiar from quantum theory 
applies) is implemented by a {\it representation}, a map $\pi$ from $\As$ into 
the set of (densely defined) operators acting on $\Hs$ which satisfies 
$\pi(cA + B) = c\pi(A) + \pi(B)$, $\pi(AB) = \pi(A)\pi(B)$ and 
$\pi(A^*) = \pi(A)^*$, for all $A,B \in \As$ and $c \in \CC$. If one finds a 
{\it state} $\omega$ on $\As$ (a linear map $\omega : \As \rightarrow \CC$
such that $\omega(A^* A) \geq 0$ for all $A \in \As$ and $\omega(I) = 1$,
where $I$ is the unit in $\As$), then such a Hilbert space $\Hs$ and 
representation $\pi$, called the GNS representation associated with $\omega$, 
can be canonically constructed from the data $(\As,\omega)$. In this space 
$\Hs$ there is a unit vector $\Omega$ such that 
$\omega(A) = < \Omega,\pi(A) \Omega>$ for all $A \in \As$. 

     In perturbative AQFT (as in some forms of deformation quantization
\cite{Wa}) the basic objects are understood as formal power series. Consider
the set $\CC[[t]]$ of all complex sequences $\{ c_0,c_1,\ldots \}$,
where to each such sequence corresponds a formal power series in the
formal parameter $t$
$$c = \sum_{n = 0}^\infty \, c_n \, t^n \, .$$
On this set addition is defined termwise and a product is defined using
Cauchy's formula:
$$ab = \big( \sum_{n = 0}^\infty \, a_n \, t^n \big) \, 
\big( \sum_{n = 0}^\infty \, b_n \, t^n \big) \equiv
\sum_{n = 0}^\infty \, \big( \sum_{m = 0}^n \, a_m b_{n-m} \big) \, t^n \, .$$
With these operations, $\CC[[t]]$ is an associative, commutative ring with unit.
If $\As$ is a complex algebra, then $\As[[t]]$ is defined as the set of
all sequences with entries in $\As$ and, again, may be thought of as the set of 
all formal power series with coefficients in $\As$. With addition, 
multiplication and scalar multiplication (with scalars in $\CC[[t]]$) defined 
similarly, $\As[[t]]$ is an algebra over $\CC[[t]]$. 

     Perturbative AQFT is based upon the causal perturbation theory developed
primarily by St\"uckelberg, Bogoliubov, Shirkov, Epstein and Glaser. This
theory cannot be described here (\cf \eg \cite{EpGl}), but it is a
mathematically rigorous renormalization theory for quantum field theoretical
perturbation series which preserves and utilizes in an essential manner
``locality'' properties, in the double sense commonly understood in AQFT ---
localization properties coupled with Einstein causality. In the papers
discussed below the ultraviolet problems with the formal series representations
of the observables are handled by techniques from causal perturbation theory.
These techniques are not described here.

     In the Fock space of the free scalar massive Bose field, consider the
cutoff interaction Hamiltonian 
$$H_I(t) = - \int g(t,{\bf x}) \, A(t,{\bf x}) \; d^3 x \, ,$$
with $g$ an infinitely differentiable function with compact support and
$A$ a (derivative of a) Wick polynomial. The corresponding time evolution
operator from time $-\tau$ to time $\tau$, where $\tau > 0$ is sufficiently
large that $(-\tau,\tau) \times \RR^3$ contains the support of $g$, is
formally given by the Dyson series
\begin{equation}  \label{dyson}
S(g) = 1 + \sum_{n = 1}^\infty \, \frac{i^n}{n!} 
\int T(A(x_1) \cdots A(x_n)) g(x_1) \cdots g(x_n) \, dx_1 \cdots dx_n \, ,
\end{equation}
where the time ordered products $T(A(x_1) \cdots A(x_n))$ are operator valued
distributions on $\Ds$ satisfying
$$T(A(x_1) \cdots A(x_n)) = T(A(x_1) \cdots A(x_k)) \, T(A(x_{k+1}) \cdots A(x_n))
$$
whenever all of $x_{k+1},\ldots,x_n$ do not lie in the forward lightcone of
any of the points $x_{1},\ldots,x_k$. Supplying the test function $g$ with
a factor (coupling constant) $\lambda$, the local S--matrix (\ref{dyson}) is 
to be understood as an element of $\As[[ \lambda ]]$, where the algebra
$\As$ is a natural extension of the Borchers class of the free field.
$S(g)$ has an inverse in $\As[[ \lambda ]]$ of the form (\ref{dyson}) with
$i$ replaced by $-i$ and the time ordered products are replaced by 
``antichronological'' products
$$\overline{T}(A(x_1) \cdots A(x_n)) \equiv 
\sum_{P \in \Ps(\{ 1,\ldots,n \})} (-1)^{ \vert P \vert + n} \prod_{p \in P}
T(A(x_i),i \in P) \, ,$$
where $\Ps(\{ 1,\ldots,n \})$ is the set of all ordered partitions of
$\{ 1,\ldots,n \}$ and $\vert P \vert$ is the number of subsets in $P$.
The ``antichronological'' products satisfy anticausal factorization.
As observed by Il'in and Slavnov \cite{IlSl}, the local S--matrices
satisfy
\begin{equation}  \label{caus}
S(f + g + h) = S(f + g)S(g)^{-1}S(g + h) \, ,
\end{equation}
whenever the support of $h$ is disjoint from the causal future of the support 
of $f$ (independently of $g$). Further solutions of (\ref{caus}) are obtained
by introducing the relative S--matrices
$$S_g(f) \equiv S(g)^{-1}S(g + f) \, ,$$
which also satisfy local commutation relations $[ S_g(h),S_g(f)] = 0$
whenever the supports of $h$ and $f$ are spacelike separated. 

     Local algebras of observables are then introduced by defining
$\As_g(\Os)$ to be the *--algebra generated by the relative S--matrices
$S_g(h)$ whose test functions $h$ have support contained in $\Os$.
If $g = g'$ in a neighborhood of a causally closed region containing $\Os$,
then there exists a unitary $V \in \As[[\lambda]]$ such that 
$V S_g(h) V^{-1} = S_{g'}(h)$ for all test functions with support in $\Os$. 
Capitalizing upon this fact, D\"utsch and Fredenhagen \cite{DuFr2} construct 
for a given interaction Lagrangian $\Ls$ (here a polynomial in the field or 
derivatives of the field) a net of such local *--algebras which 
satisfies Einstein causality and is covariant under a natural adjoint action 
of the usual unitary representation of the Poincar\'e group defined on the 
initial Fock space.

     D\"utsch and Fredenhagen \cite{DuFr1} have shown that this approach
can be employed also for QED. Because the basic quantities are local and
independent of the test function $g$ in the sense indicated above, no
reference need be made to the adiabatic limit. The incompatibility in QED
between gauge invariance, locality and positive definite inner products on the
state space when the unobservable gauge potentials and charge carrying fields
are employed is addressed by a local construction of the 
observables and of the physical Hilbert space in which the observables are
faithfully represented (once again as formal power series of unbounded 
operators).

     More recently, Brunetti, D\"utsch and Fredenhagen \cite{BrDuFr} have
refined this approach to allow the treatment of low dimensional theories and
non-polynomial interactions. In addition, they studied in this framework
three of the various approaches to renormalization group ideas that are
in use among theoretical physicists and have established their mutual logical 
relations. They obtain an algebraic form of the Callan--Symanzik equation and 
compute the $\beta$ function in the $\phi_4^4$ and $\phi_6^3$ interactions in 
their framework, finding perfect agreement between their results and those 
found by heuristic methods.

     Although this work does not provide quantum field models satisfying the
HAK or Wightman axioms, it falls comfortably within the framework of AQFT,
since the primary objects are, again, nets of local *--algebras generated
by observables which are Poincar\'e covariant and satisfy Einstein causality
and, again, the work is carried out with complete mathematical rigor. 
However, when these authors speak of representations in Hilbert spaces,
the Hilbert spaces are vector spaces over the field $\CC[[\lambda]]$, not
over $\CC$. So taking expectations of observables in states in this
approach results in a formal complex power series, not a complex number.
Hence, in order to make the connection to experiments one must deliberately
consider a partial sum of this series, \ie consider the perturbation series
only to a finite order, as is done in heuristic QFT. Since these series are
not convergent, one is returned to the question ``Is there an exact model?''   

\section{Outlook} \label{outlook}
\setcounter{equation}{0}

     It is evident that the efforts of the constructive quantum field
theorists have been crowned with many successes. They have constructed 
superrenormalizable models, renormalizable models and even nonrenormalizable
models, as well as models which fall outside of that classification scheme
since they apparently do not correspond to some classical Lagrangian.
And they have found means to extract rigorously from these models physically 
and mathematically crucial information. In many of these models the HAK and
the Wightman axioms have been verified. In the models constructed to this 
point, the intuitions/hopes of the quantum field theorists have been 
largely confirmed. However, local gauge theories such as quantum 
electrodynamics, quantum chromodynamics and the Standard Model --- precisely 
the theories whose approximations of various kinds are used in a central 
manner by elementary particle theorists and cosmologists --- remain 
to be constructed. These models present significant mathematical and conceptual 
challenges to all those who are not satisfied with {\it ad hoc} and 
essentially instrumentalist computation techniques.
     
     Why haven't these models of greatest physical interest been constructed
yet (in any mathematically rigorous sense which preserves the basic principles
constantly evoked in heuristic QFT and does not satisfy itself with an 
uncontrolled approximation)? Certainly, one can point to the practical 
fact that only a few dozen people have worked in CQFT. This should be compared 
with the many hundreds working in string theory and the thousands who have 
worked in elementary particle physics. Progress is necessarily slow if only a 
few are working on extremely difficult problems.\footnote{Indeed, many workers 
in CQFT have chosen to take the methods developed for the purposes of CQFT and 
to apply them instead to problems in statistical mechanics, many body physics 
and solid state physics, where progress has been much easier.} It may well be 
that patiently proceeding along the lines indicated above and steadily 
improving the technical tools employed will ultimately yield the desired 
rigorous constructions. It may also be the case that a completely new approach 
is required, though remaining within the CQFT program as described in 
Section \ref{intro}, something whose essential novelty is analogous to the 
differences between the approaches in Section \ref{algebra1}, \ref{path1},
\ref{stoch} and \ref{algebra2}. 

     It may even be the case that, as Gurau, Magnen and Rivasseau have written 
\cite{GuMaRi}, ``perhaps axiomatization of QFT might have been premature''; in 
other words, perhaps the Wightman and HAK axioms do not provide the proper 
mathematical framework for QED, QCD, SM, even though, as the constructive 
quantum field theorists have so convincingly demonstrated, that framework is 
quite suitable for so many models of such varying types and, as the algebraic 
quantum field theorists have just as convincingly demonstrated, that framework 
is flexible and powerful when dealing with the conceptual and mathematical 
problems in QFT which go {\it beyond} mathematical existence. But it is
possible that the mathematically and conceptually essential core of a rigorous 
formulation of QFT that can include the missing models lies somewhere else. 
Certainly, there are presently many attempts to 
understand aspects of QFT from the perspective of mathematical ideas which are 
quite unexpected when seen from the vantage point of current QFT and even 
from the vantage point of quantum theory itself, as rigorously formulated
by von Neumann and many others. These speculations, as suggestive as some may 
be, are currently beyond the scope of this article. 


\end{document}